\documentstyle[amsmath,amssymb,12pt,epsfig]{article}
\begin{document}
\setlength{\parskip}{0.45cm}
\setlength{\baselineskip}{0.75cm}
%
%
%
\begin{titlepage}
\setlength{\parskip}{0.25cm}
\setlength{\baselineskip}{0.25cm}
\begin{flushright}
DO-TH 2001/03\\
\vspace{0.2cm}
February 2001
\end{flushright}
\vspace{1.0cm}
\begin{center}
\LARGE
{\bf Spin-Dependent Structure Functions}\\
\LARGE{\bf of Real and Virtual Photons}
\vspace{1.5cm}

\large
M. Gl\"uck, E.\ Reya, C.\ Sieg\\
\vspace{1.0cm}

\normalsize
{\it Universit\"{a}t Dortmund, Institut f\"{u}r Physik,}\\
{\it D-44221 Dortmund, Germany} \\
\vspace{0.5cm}

\vspace{1.5cm}
\end{center}

\begin{abstract}
\noindent
The implications of the positivity constraint, $|g_1^{\gamma(P^2)}(x,Q^2)|
\leq F_1^{\gamma(P^2)}(x,Q^2)$, on the presently unknown spin--dependent
structure function $g_1^{\gamma(P^2)}(x,Q^2)$ of real and virtual photons
are studied at scales $Q^2\gg P^2$ where longitudinally polarized photons
dominate physically relevant cross sections.  In particular it is shown
how to implement the physical constraints of positivity and continuity at
$P^2=0$ in NLO calculations which afford a nontrivial choice of suitable
(DIS) factorization schemes related to $g_1^{\gamma}$ and $F_1^{\gamma}$
and appropriate boundary conditions for the polarized parton distributions
of real and virtual photons.  The predictions of two extreme  `maximal'
and   `minimal' saturation scenarios are presented and compared with results obtained
within the framework of a simple quark   `box' calculation expected to 
yield reasonable estimates in the not too small regions of $x$ and $P^2$.
\end{abstract}
\end{titlepage}


\section{Introduction}

The structure functions of photons with momentum $p$ and virtuality 
$P^2=-p^2$ probed at a scale 
$Q^2$ \raisebox{-0.1cm}{$\stackrel{>}{\sim}$} 1 GeV$^2$ 
{\underline{and}} $Q^2\gg P^2$, i.e. in the  `Bjorken limit', can be
described in terms of photonic parton distributions.  These 
spin--[in]dependent parton distributions  of $\gamma(P^2)$,
henceforth denoted by $\left[f^{\gamma(P^2)}(x,Q^2)\right]$ and
$\Delta f^{\gamma(P^2)}(x,Q^2)$ with $f=q,\bar{q},g$ and $q=u,d,s$,
provide the dominant, lowest--twist, contributions to the structure
functions $\left[F_{1,2}^{\gamma(P^2)}(x,Q^2)\right]$ and 
$g_1^{\gamma(P^2)}(x,Q^2)$ in [un]polarized deep elastic $ep$ collisions.

One expects, of course, a unified description of real ($P^2=0$) and 
virtual ($P^2\neq 0$) photons in the sense of continuity of all the
physical predictions at $P^2=0$.  It turns out, however, that the 
so--called `direct' contribution to the structure functions due to 
the subprocess $\gamma^*(Q^2)\gamma(P^2)\to q\bar{q}$ which arises
at the next--to--leading order (NLO) analysis of $F_{1,2}^{\gamma(P^2)}$
and $g_1^{\gamma(P^2)}$ is {\underline{discontinuous}} at $P^2=0$, 
thus violating the basic continuity demand.

It was pointed out in \cite{ref1} that the physically compelling
continuity at $P^2=0$ is facilitated by treating the direct--photon
contribution at $P^2\neq 0$ to be the {\underline{same}} as for a 
real on--shell ($P^2=0$) photon.  As a logical consequence of this 
approach to the continuity demand it is mandatory to consider the 
direct contribution of the virtual photon to {\underline{any}} deep
inelastic scattering (DIS) process as if it was {\underline{real}}.
In \cite{ref1} the consequences of this approach to spin--independent
DIS processes were presented.  The present paper extends this study
to the spin--dependent DIS processes.

In Section 2 we present some consequences of our unified approach as
reflected by the spin--dependent structure function $g_1^{\gamma(P^2)}
(x,Q^2)$ characterizing the spin--dependent deep--inelastic inclusive
$\gamma^*(Q^2)\gamma(P^2)\to$ hadrons scattering process accessible 
in longitudinally polarized $e^+e^-$ annihilations.  Here the various
possible (input) boundary conditions for the polarized parton
distributions and structure functions of longitudinally polarized real
and virtual photons are discussed and presented together with their 
formal analytic solutions of the inhomogeneous renormalization group
(RG) $Q^2$--evolution equations.  Various illustrative quantitative
expectations are presented in Section 3.  These QCD resummed RG--improved
calculations are compared in Section 4 with the predictions of the
standard non--resummed  `naive' quark--parton model (QPM) QED  `box'
approach.  In particular, the relevance of the polarized gluonic photon
content, the typical RG--improved QCD ingredient, is studied within 
this context.  Finally, our conclusions are summarized in Section 5.

\renewcommand{\theequation}{\arabic{section}.\arabic{equation}}
\section{$g_1^{\gamma(P^2)}(x,Q^2)$ and the associated parton 
 distributions of polarized real and virtual photons}

The flux of (longitudinally) polarized virtual photons produced by the 
the bremsstrahlung process of high energy electrons $e(k)\to e(k')
+\gamma (p)$, $P^2\equiv -p^2=-(k'-k)^2$, at $e^+e^-$ or $ep$ colliders
is given by \cite{ref2}
\begin{equation}
\Delta f_{\gamma(P^2)/e}(y) = \frac{\alpha}{2\pi}\,
  \left[ \frac{1-(1-y)^2}{y}\, \frac{1}{P^2} 
     + \frac{2m_e^2y^2}{P^4} \right] 
\end{equation}
where $y=E_{\gamma}/E_e$ and $\alpha\simeq 1/137$.  The real ($P\simeq 0$)
photons usually considered are those whose virtuality is in reality 
very small, i.e.\ of order $P_{\rm min}^2={\cal{O}}(m_e^2)$ or, 
experimentally, at least $P^2<10^{-2}$ GeV$^2$ which is the case for
the bulk of produced photons in untagged or antitagging experiments.
On the other hand, a sizeable finite virtuality is achieved by tagging
of the outgoing electron at the photon producing vertex $e\to e\gamma$.
Whenever these virtual photons, with their virtuality being 
{\underline{entirely}} taken care of by the {\underline{flux}} factor
in (2.1), are probed at a scale $Q^2\gg P^2$ they may be considered 
as {\underline{real}} photons, which means that cross sections of partonic
subprocesses involving $\gamma(P^2)$ should be calculated as if $P^2=0$
(partly due to the suppression of any additional terms).  Furthermore,
the polarized parton distributions $\Delta f^{\gamma(P^2)}(x,Q^2)$ of
the virtual photon obey the same $Q^2$ evolution, i.e. renormalization
group (RG), equations as the real photon $\gamma\equiv\gamma(P^2=0)$ 
distributions $\Delta f^{\gamma}(x,Q^2)$ and the only difference between
them resides in the different boundary conditions.  This concept is
similar to the one suggested and developed in \cite{ref1} for unpolarized
photons which, as emphasized in the Introduction, follows from the basic
continuity demand at $P^2=0$.

Following the situation of protons \cite{ref3} and unpolarized photons
\cite{ref1}, the structure function of polarized photons $g_1^{\gamma(P^2)}
(x,Q^2)$ will be decomposed into $g_{1,\ell}^{\gamma(P^2)}$ due to the
light (massless) $u,d,s$ partons, and $g_{1,h}^{\gamma(P^2)}$ due to the
heavy $h=c,b,\ldots$ quarks whose contributions are calculated in fixed
order of perturbation theory which are known and unproblematic due to 
the finite $m_h\neq 0$ (we shall come back to this point at the end of
this Section).  The main issue of this paper resides of course in 
$g_{1,\ell}^{\gamma(P^2)}$ which, up to NLO($\overline{\rm MS}$), is 
given by the following expression: 
\begin{eqnarray}
g_{1,\ell}^{\gamma(P^2)}(x,Q^2) & = & \frac{1}{2} \sum_{q=u,d,s}
  e_q^2\left\{ \Delta q^{\gamma(P^2)}(x,Q^2) 
      + \Delta\bar{q}\,^{\gamma(P^2)}(x,Q^2)
        + \frac{\alpha_s(Q^2)}{2\pi}\right.\nonumber\\
& & \left. 
 \times\left[ \Delta C_q\otimes\Delta (q+\bar{q})^{\gamma(P^2)}
         + 2 \Delta C_g\otimes \Delta g^{\gamma(P^2)}\right]
        + e_q^2\, \frac{\alpha}{\pi}\, \Delta C_{\gamma}(x)\right\}
\end{eqnarray}
where $\otimes$ denotes the usual convolution integral.  Here,
$\Delta\bar{q}\,^{\gamma(P^2)}=\Delta q^{\gamma(P^2)}$ and 
$\Delta g^{\gamma(P^2)}$ provide the so--called   `resolved' contribution
of $\gamma(P^2)$ to $g_{1,\ell}^{\gamma(P^2)}$ with the usual hadronic
polarized Wilson coefficient functions in the conventional 
$\overline{\rm MS}$ factorization scheme given by \cite{ref4,ref5}
\begin{eqnarray}
\Delta C_q(x) & = & \frac{4}{3} \left[ (1+x^2)
    \left(\frac{\ln(1-x)}{1-x}\right)_+
     -\frac{3}{2}\, \frac{1}{(1-x)}_+ -\frac{1+x^2}{1-x}\,\ln x + 2+x\right.
\nonumber\\
& &       
\left.-\left( \frac{9}{2}+\frac{\pi^2}{3}\right) \delta(1-x)\right]
\nonumber\\
\Delta C_g(x) & = & \frac{1}{2}
     \left[(2x-1)\left(\ln \frac{1-x}{x}-1\right)
       +2(1-x)\right]\, .
\end{eqnarray}
The aforementioned `direct' contribution is provided by the 
$\Delta C_\gamma(x)$ term in (2.2) which has to be calculated for
{\underline{real}} photons $\gamma\equiv\gamma(P^2=0)$, as follows from
the continuity condition, in the polarized  `box' subprocess
$\gamma^*(Q^2)\gamma\to q\bar{q}$.  Thus $\Delta C_{\gamma}$ can be
easily obtained from $\Delta C_g$ in (2.3) which is also derived for
a massless on--shell gluon in the polarized subprocess $\gamma^*(Q^2)g
\to q\bar{q}$: 
\begin{equation}
\Delta C_{\gamma}(x) = \frac{3}{(1/2)}\Delta C_g(x) =
  3\left[ (2x-1)\left(\ln\frac{1-x}{x}-1\right) +2(1-x)\right] \, .
\end{equation}

The NLO coefficient functions $\Delta C_{q,g,\gamma}$ are obviously
factorization scheme dependent and we shall follow the traditional
choice \cite{ref6}, motivated by the perturbative stability of 
unpolarized photon structure functions, where $\Delta C_{q,g}$ are
considered in the $\overline{\rm MS}$ scheme while the destabilizing
$\Delta C_{\gamma}$ term in (2.2), as given by (2.4), is entirely
absorbed \cite{ref1,ref6,ref7} into the $\overline{\rm MS}$ (anti)quark
densities in (2.2) as implied by the so-called `polarized DIS$_{\gamma}$'
factorization scheme, to be denoted by DIS$_{\Delta\gamma}$:
\begin{eqnarray}
(\Delta q+\Delta\bar{q})_{\rm DIS_{\Delta\gamma}}^{\gamma(P^2)} & = &
   (\Delta q+\Delta\bar{q})^{\gamma(P^2)} + e_q^2\,\frac{\alpha}{\pi}\,
      \Delta C_{\gamma}(x)\nonumber\\
\Delta g_{\rm DIS_{\Delta\gamma}}^{\gamma(P^2)} & = & \Delta g^{\gamma(P^2)}\, .
\end{eqnarray}
This redefinition of parton distributions implies that the polarized
NLO($\overline{\rm MS}$) splitting functions $\Delta k_{q,g}^{(1)}(x)$ 
of the photon into quarks and gluons, appearing in inhomogeneous 
NLO RG $Q^2$--evolution equations \cite{ref7} 
for $\Delta f^{\gamma(P^2)}(x,Q^2)$, have correspondingly to be transformed 
according to \cite{ref6,ref8}
\begin{eqnarray}
\Delta k_q^{(1)}|_{\rm DIS_{\Delta\gamma}} & = & 
   \Delta k_q^{(1)} - e_q^2\, \Delta P_{qq}^{(0)}
       \otimes \Delta C_{\gamma}\nonumber\\
\Delta k_g^{(1)}|_{\rm DIS_{\Delta\gamma}} & = & 
    \Delta k_g^{(1)} - 2 \sum_q e_q^2\, \Delta P_{gq}^{(0)}
        \otimes \Delta C_{\gamma} 
\end{eqnarray}
\vspace{-0.5cm}
where \cite{ref7}
\begin{eqnarray}
\Delta k_q^{(1)}(x) & = &\frac{1}{2}\, 3 e_q^2\, \frac{4}{3}
   \biggl\{ -9\ln x + 8(1-x)\,\ln (1-x) + 27x -22\nonumber\\
& & +  (2x-1)\left[ \ln^2 x + 2 \ln^2(1-x) -4\ln x\ln(1-x)
        -\frac{2}{3}\pi^2\right] \biggr\}\nonumber\\
\Delta k_g^{(1)}(x) & = & 3\sum_q e_q^2\,\frac{4}{3}
    \left\{ -2(1+x)\, \ln^2 x+2(x-5)\, \ln x -10(1-x)\right\} 
\end{eqnarray}
with $\Delta k_q^{(1)}$ referring to each single (anti)quark flavor.  
The polarized LO splitting functions are given by 
$\Delta P_{qq}^{(0)}=\frac{4}{3}\left(\frac{1+x^2}{1-x}\right)_+$ 
and $\Delta P_{gq}^{(0)}=\frac{4}{3}(2-x)$. The NLO expression for 
$g_{1,\ell}^{\gamma(P^2)}$ in the  DIS$_{\Delta\gamma}$ scheme is thus 
given by (2.2) with $\Delta C_{\gamma}(x)$ being {\underline{dropped}}.
Since from now on we shall exclusively work in this DIS$_{\gamma}$
scheme, we skip the label   `DIS$_{\Delta\gamma}$' on all our 
subsequent parton distributions and splitting functions.  The leading
order (LO) expression for $g_{1,\ell}^{\gamma(P^2)}$ is obviously
obtained from eq.\ (2.2) by simply setting $\Delta C_{q,g,\gamma}=0$.

The general solution of the {\underline{inhomogeneous}} evolution
equations \cite{ref7} for $\Delta f^{\gamma(P^2)}(x,Q^2)$ may be
written as 
\begin{equation}
\Delta f^{\gamma(P^2)}(x,Q^2) = \Delta f_{p\ell}^{\gamma(P^2)}(x,Q^2)
  +\Delta f_{\rm had}^{\gamma(P^2)}(x,Q^2)
\end{equation}
and similarly for $g_{1,\ell}^{\gamma(P^2)}(x,Q^2)$ in (2.2).
The nonhadronic `pointlike' component $\Delta f_{p\ell}^{\gamma(P^2)}$
evolves according to the full {\underline{in}}homogeneous evolution
equations subject to the boundary condition 
\begin{equation}
\Delta f_{p\ell}^{\gamma(P^2)}(x,\tilde{P}^2) = 0\, , \quad \quad
   \tilde{P}^2= {\rm max}\, (P^2,\mu^2)
\end{equation}
with $\mu$ being some appropriately chosen resolution scale taken
here, in the spirit of the radiative parton model \cite{ref3}, to
be \cite{ref1,ref3} $\mu_{\rm NLO(LO)}^2 = 0.40$ (0.26) GeV$^2$.
The `hadronic' component in (2.8) represents the solution to the
conventional {\underline{homogeneous}} evolution equations where
the photon splitting functions $\Delta k_{q,g}^{(0,1)}$ are dropped.
  Following refs.\
\cite{ref7,ref9} we shall study two extreme scenarios for
$\Delta f_{\rm had}^{\gamma(P^2)}(x,Q^2)$:

\noindent (i)$\quad$a  `maximal' scenario corresponding to a NLO
input
\begin{equation}
\Delta f_{\rm had}^{\gamma(P^2)}(x,\tilde{P}^2)) = \eta(P^2)\, 
      f_{\rm had}^{\gamma}(x,\tilde{P}^2)_{\rm DIS_{\gamma,1}}
\end{equation}
where $\eta(P^2)=(1+P^2/m_{\rho}^2)^{-2}$ is a dipole suppression
factor with $m_{\rho}^2=0.59$ GeV$^2$, and in LO the unpolarized
photonic parton distributions 
$f_{\rm had}^{\gamma}(x,\tilde{P}^2)_{\rm LO}$ refer to the 
common LO (input) densities as obtained, for example, in
\cite{ref1};

\noindent (ii)$\quad$a  `minimal' scenario corresponding to a NLO
input \cite{ref10} 
\begin{eqnarray}
\Delta q_{\rm had}^{\gamma(P^2)}(x,\tilde{P}^2) & = & 
  \eta(P^2)e_q^2 \, \frac{\alpha}{2\pi} \left[C_{\gamma,1}(x)
    - C_{\gamma,2}(x)\right]\nonumber\\
&  = & \eta(P^2) e_q^2\, \frac{\alpha}{2\pi} 
         \left[-12x(1-x)\right]\nonumber\\
\Delta g_{\rm had}^{\gamma(P^2)}(x,\tilde{P}^2) & = & 0\, ,
\end{eqnarray}
whereas in LO $\Delta q_{\rm had}^{\gamma}(x,\tilde{P}^2)$ also
vanishes, due to the absence of the (unpolarized) NLO coefficient
functions $C_{\gamma,i}$, and thus the input 
$\Delta f^{\gamma(P^2)}(x,\tilde{P}^2)_{\rm LO}=0$ coincides
with the  `pointlike' LO solution for $\Delta f^{\gamma(P^2)}
(x,Q^2)_{\rm LO}$ in (2.8) with \cite{ref1} 
$\mu_{\rm LO}^2 = 0.26$ GeV$^2$.  Thus this  `minimal' scenario
constitutes a genuinely lowest, but possibly unrealistic, limit
for the expected parton distributions of a longitudinally polarized
photon.  

It should be noticed that the unpolarized NLO inputs in (2.10) 
and (2.11) refer to the so--called  DIS$_{\gamma,1}$ factorization
scheme \cite{ref10} related to $F_1^{\gamma}$, rather than to
$F_2^{\gamma}$, in order to comply with the fundamental
positivity constraint $|A_1^{\gamma(P^2)}|\equiv |g_1^{\gamma(P^2)}
/F_1^{\gamma(P^2)}|\leq 1$, to which we shall turn in more detail
in the next Section.  The unpolarized parton distributions in the
DIS$_{\gamma,1}$ factorization scheme required in (2.10) can be
easily derived from the ones in the DIS$_{\gamma}$ scheme
\cite{ref1,ref6,ref8}, as obtained from an analysis \cite{ref1}
of the data on $F_2^{\gamma}$, via \cite{ref10}
\begin{eqnarray}
q^{\gamma}(x,Q^2)_{\rm DIS_{\gamma,1}} & = &
   q^{\gamma}(x,Q^2)_{\rm DIS_{\gamma}} -e_q^2 \,\frac{\alpha}{2\pi}\,
        12x(1-x)\nonumber\\
g^{\gamma}(x,Q^2)_{\rm DIS_{\gamma,1}}& = & 
   g^{\gamma}(x,Q^2)_{\rm DIS_{\gamma}}\,\,  .
\end{eqnarray}
These boundary conditions are dictated, as in \cite{ref1}, by
the continuity of $\Delta f^{\gamma(P^2)}(x,Q^2)$ at $P^2=0$.
The hadronic vector--meson--dominance (VMD) oriented input 
distributions of the unpolarized real photon $f_{\rm had}^{\gamma}
(x,Q^2)$ in (2.10) will also be taken from \cite{ref1} for reasons
of consistency with the positivity constraint.  These scenarios
will be considered below for our quantitative analyses.

It should be mentioned that our above boundary conditions for
the hadronic input in NLO differ substantially from those 
considered by Sasaki and Uematsu \cite{ref11,ref12} who consider
only the kinematical region $\Lambda^2\ll P^2\ll Q^2$ in contrast
to our analysis addressing the full kinematical region $0\leq P^2
\ll Q^2$ and the ensuing continuity constraints at $P^2=0$.
More specifically, Sasaki and Uematsu \cite{ref11,ref12} consider
on the contrary the perturbatively calculable doubly--virtual
polarized box $\gamma^*(Q^2)\gamma(P^2)\to q\bar{q}$, following
the original treatment of the parton structure of the unpolarized
virtual photon \cite{ref13}, and thus adopt for 
$\Delta f_{\rm had,NLO}^{\gamma(P2)}$ the following boundary
condition at $Q^2=P^2$ in the DIS$_{\gamma}$ factorization
scheme:   
\begin{eqnarray}
\Delta q_{\rm had,NLO}^{\gamma(P^2)}(x,Q^2=P^2) & = &
  \Delta\bar{q}_{\rm had,NLO}^{\gamma(P2)}(x,Q^2=P^2)\nonumber\\
& = & 3e_q^2\, \frac{\alpha}{2\pi}\,(2x-1)
    \left(\ln \frac{1}{x^2}-2\right)\nonumber\\
\Delta g_{\rm had,NLO}^{\gamma(P^2)}(x,Q^2=P^2) & = & 0\, .
\end{eqnarray}
Due to the nonvanishing virtuality $(P^2\neq 0)$ of the target
photon, these results obviously cannot be related anymore, as in
(2.4), to the one of a massless initial on--shell gluon in (2.3).
Apart from exluding the polarized real $(P^2=0)$ photon within
this approach, the input in (2.13) is problematic on its own since
kinematically $x$ is restrained to $x\leq \frac{1}{2}$ at
$Q^2=P^2$ due to $0\leq x \leq (1+P^2/Q^2)^{-1}$.  This latter
problem is also faced in the treatment of the partonic structure
of a virtual unpolarized photon as suggested originally in
\cite{ref14}.  (The LO boundary conditions are, in contrast to
(2.13), obviously given by 
\mbox{$\Delta f_{\rm had,LO}^{\gamma(P^2)}(x,Q^2=P^2)=0$.)}
In view of $P^2\gg \Lambda^2$ it is tacitly assumed in 
\cite{ref11,ref12} that the hadronic VMD input, eq.\ (2.10), is
negligible.  One can, however, also implement a smooth transition
to $P^2=0$ in this approach by multiplying the r.h.s.\ of (2.13)
by, say \cite{ref15}, $\zeta(P^2)=P^2/(P^2+Q_0^2)$ where 
$Q_0^2\simeq 1$ GeV$^2$, as derived from DIS $ep$ structure 
functions, and {\underline{adding}} to this part also the VMD
hadronic component in (2.10).  Alternatively a smooth transition
to $P^2=0$ may be achieved by multiplying the r.h.s. of (2.13)
by $[1-\eta(P^2)]$ with $\eta(P^2)$ as in (2.10), as has been
originally suggested for the unpolarized virtual photon \cite{ref14}.

Having fixed the boundary (input) conditions, we turn now to the
inhomogeneous RG $Q^2$--evolution equations which are formally 
very similar to the ones of an unpolarized real photon \cite{ref6,ref8},
replacing the spin--independent splitting functions everywhere by 
their spin--dependent counterparts \cite{ref7}.  Their LO and NLO
solutions can be given analytically for the Mellin $n$--moments of
$\Delta f^{\gamma(P^2)}(x,Q^2)$ in (2.8):
\begin{eqnarray}
\Delta f^{\gamma(P^2),n}(Q^2) & \equiv &
  \int_0^1 dx\, x^{n-1} \Delta f^{\gamma(P^2)}(x,Q^2)\nonumber\\
& = & \Delta f_{p\ell}^{\gamma(P^2),n}(Q^2) +
        \Delta f_{\rm had}^{\gamma(P^2),n}(Q^2)\, .
\end{eqnarray}

The  `pointlike' solution, which vanishes at the input scale
$Q^2=\tilde{P}^2$ in (2.9), is driven by the LO and NLO pointlike
photon splitting functions $\Delta k_{q,g}^{(0,1)}$ appearing in
the inhomogeneous evolution equations, while 
$\Delta f_{\rm had}^{\gamma(P^2)}$ depends on the hadronic input
in (2.10) and evolves according to the standard homogeneous 
evolution equations.  The flavor--singlet solutions for $f=3$
flavors, i.e. $Q^2\equiv Q_3^2\leq m_c^2$, are given by
\begin{eqnarray}
\renewcommand{\arraystretch}{2.0}
\left( \begin{array}{c}
\Delta \Sigma_{p\ell}^{\gamma(P^2),n}(Q_3^2)\\
\Delta g_{p\ell}^{\gamma(P^2),n}(Q_3^2) 
\end{array} \right)
& = & 
 \frac{4\pi}{\alpha_s^{(3)}(Q_3^2)} 
  \left(1+\frac{\alpha_s^{(3)}(Q_3^2)}{2\pi}\Delta\hat{U}\right)
   \left[ 1-L_3^{1-\frac{2}{\beta_0^{(3)}}
    \Delta\hat{P}^{(0)n}}\right]
\nonumber\\
& &  
\times
\frac{1}{1-\frac{2}{\beta_0^{(3)}}\Delta\hat{P}^{(0)n}}\,
 \frac{\alpha}{2\pi\beta_0^{(3)}}\Delta\vec{k}\,^{(0)n} +
  \left[1-L_3^{-\frac{\large 2}{\beta_0^{(3)}}\Delta\hat{P}^{(0)n}}\right]
\nonumber\\
& & 
\times
 \frac{1}{-\Delta\hat{P}^{(0)n}}\frac{\alpha}{2\pi}
  \left(\Delta\vec{k}^{(1)n} -\frac{\beta_1^{(3)}}{2\beta_0^{(3)}}
   \Delta\vec{k}^{(0)n}-\Delta\hat{U}\Delta\vec{k}^{(0)n}\right)
\end{eqnarray}

\begin{eqnarray}
\renewcommand{\arraystretch}{2.0}
\left( \begin{array}{c}
\Delta \Sigma_{\rm had}^{\gamma(P^2),n}(Q_3^2)\\
\Delta g_{\rm had }^{\gamma(P^2),n}(Q_3^2) 
\end{array} \right)
& = & 
 \left[ L_3^{-\frac{2}{\beta_0^{(3)}}\Delta\hat{P}^{(0)n}}
  +\,\frac{\alpha_s^{(3)}(Q_3^2)}{2\pi}\Delta\hat{U}
    L_3^{-\frac{2}{\beta_0^{(3)}}\Delta\hat{P}^{(0)n}}\right.
\nonumber\\
& & 
  \left. -\frac{\alpha_s^{(3)}(\tilde{P}^2)}{2\pi}
   L_3^{-\frac{2}{\beta_0^{(3)}}\Delta\hat{P}^{(0)n}}
    \Delta\hat{U}\right]
\left( \begin{array}{c}
\Delta \Sigma_{\rm had}^{\gamma(P^2),n}(\tilde{P}^2)\\
\Delta g_{\rm had }^{\gamma(P^2),n}(\tilde{P}^2) 
\end{array} \right)
\end{eqnarray}
where $\Delta\Sigma = \Sigma_f(\Delta q+\Delta\bar{q})$,
$\,\Delta\vec{k}\,^{(0)}=(\Delta k_{\Sigma}^{(0)},0)^T$ and
$\Delta\vec{k}\,^{(1)}=(\Delta k_{\Sigma}^{(1)},\, 
\Delta k_g^{(1)})^T$ with 
$\Delta k_{\Sigma}^{(0,1)}=2\Sigma_q\Delta k_q^{(0,1)}$
denote the inhomogeneous LO and NLO polarized photon splitting 
functions into quarks and gluons \cite{ref7} in (2.6) and
$\Delta k_q^{(0)}=\frac{1}{2}\,3e_q^2\, 2[2x-1]$.  The $2\times 2$
matrix $\Delta\hat{U}$ is, in complete analogy to the unpolarized
case \cite{ref6}, expressed in terms of the usual $2\times 2$
matrices of the polarized one-- and two--loop splitting functions
$\Delta\hat{P}^{(0)n}$ and $\Delta\hat{P}^{(1)n}$ which have
been presented in \cite{ref16} and from where also the $n$--moments
of the coefficient functions in (2.3) and (2.4) can be
obtained.  The input distributions 
$\Delta f_{\rm had}^{\gamma(P^2),n}(\tilde{P}^2)$ in (2.16)
are given by (2.10) and 
$L_3\equiv\alpha_s^{(f=3)}(Q_3^2)/\alpha_s^{(3)}(\tilde{P}^2)$.
For $0\leq P^2\leq \mu^2$ one of course has to freeze
$\alpha_s^{(3)}(\tilde{P}^2)$ at $\tilde{P}^2=\mu^2$ in order
 to comply with the
LO and NLO boundary conditions in (2.10).

Evoluting beyond the $\overline{\rm MS}$   `threshold' $Q_3=m_c$,
one has to take into account $f+1=4$ active flavors in 
$\alpha_s^{(f+1)}(Q^2)$ at $Q^2\equiv Q_4^2>m_4^2\equiv m_c^2$ and
where the results obtained in (2.15) and (2.16) at $Q_3^2$ serve
as input for the {\underline{hadronic}} component of the full
solution in (2.14).  In this way we arrive at the following general
form of solutions which holds, in an obvious way, for any active
number of flavors, i.e.\ $m_4<Q_4\leq m_5\equiv m_b$ as well as for
$m_5<Q_5\leq m_6\equiv m_t$:  
\begin{eqnarray}
\renewcommand{\arraystretch}{2.0}
\left( \begin{array}{c}
\Delta \Sigma_{p\ell}^{\gamma(P^2),n}(Q_{f+1}^2)\\
\Delta g_{p\ell}^{\gamma(P^2),n}(Q_{f+1}^2) 
\end{array} \right)
& = & 
 \frac{4\pi}{\alpha_s^{(f+1)}(Q_{f+1}^2)} 
   \left[1-L_{f+1}^{1-\frac{2}{\beta_0^{(f+1)}}
     \Delta\hat{P}^{(0)n}}\right]
\nonumber\\
& &  
\times
\frac{1}{1-\frac{2}{\beta_0^{(f+1)}}\Delta\hat{P}^{(0)n}}\,
 \frac{\alpha}{2\pi\beta_0^{(f+1)}}\Delta\vec{k}\,^{(0)n}+
  \left[1-L_{f+1}^{-\frac{2}{\beta_0^{(f+1)}}
    \Delta\hat{P}^{(0)n}}\right]
\nonumber\\
& & 
\times
 \frac{1}{-\Delta\hat{P}^{(0)n}}\frac{\alpha}{2\pi}
  \left(\Delta\vec{k}^{(1)n} -\frac{\beta_1^{(f+1)}}{2\beta_0^{(f+1)}}
   \Delta\vec{k}^{(0)n}-\Delta\hat{U}\Delta\vec{k}^{(0)n}\right)
\end{eqnarray}

\begin{eqnarray}
\renewcommand{\arraystretch}{2.0}
\left( \begin{array}{c}
\Delta \Sigma_{\rm had}^{\gamma(P^2),n}(Q_{f+1}^2)\\
\Delta g_{\rm had}^{\gamma(P^2),n}(Q_{f+1}^2) 
\end{array} \right)
& = & 
 \left[ L_{f+1}^{-\frac{2}{\beta_0^{(f+1)}}\Delta\hat{P}^{(0)n}}
  +\frac{\alpha_s^{(f+1)}(Q_{f+1}^2)}{2\pi}\Delta\hat{U}
    L_{f+1}^{-\frac{2}{\beta_0^{(f+1)}}\Delta\hat{P}^{(0)n}}\right.
\nonumber\\
& & 
  \left. -\,\frac{\alpha_s^{(f+1)}(m_{f+1}^2)}{2\pi}
   L_{f+1}^{-\frac{2}{\beta_0^{(f+1)}}\Delta\hat{P}^{(0)n}}
    \Delta\hat{U}\right]
\nonumber\\
& & 
\times
\left( \begin{array}{c}
\Delta\Sigma_{p\ell}^{\gamma(P^2),n}(m_{f+1}^2)  
   +\,\Delta \Sigma_{\rm had}^{\gamma(P^2),n}(m_{f+1}^2)\\
\Delta g_{p\ell}^{\gamma(P^2),n}(m_{f+1}^2) 
   +\,\Delta g_{\rm had }^{\gamma(P^2),n}(m_{f+1}^2) 
\end{array} \right)
\end{eqnarray}
where 
$L_{f+1}\equiv \alpha_s^{(f+1)}(Q_{f+1}^2)/\alpha_s^{(f+1)}(m_{f+1}^2)$
and it should be noted that the `pointlike' solution
is always such that it vanishes, per definition, at each
$Q_{f+1}=m_{f+1}$, as in (2.9) and (2.15) at $Q^2\equiv Q_3^2 =
\tilde{P}^2$. (Similar solutions have been used for calculating
the parton content of unpolarized photons \cite{ref6,ref14}).
The evolution of $\alpha_s^{(f)}(Q^2)$, corrresponding to a 
number of $f$ active flavors, is obtained by exactly solving in
NLO($\overline{\rm MS}$)
\begin{equation}
\frac{d\alpha_s^{(f)}(Q^2)}{d\ln Q^2} = -\frac{\beta_0^{(f)}}{4\pi}
 \left[\alpha_s^{(f)}(Q^2)\right]^2 - \frac{\beta_1^{(f)}}{16\pi^2}
  \left[\alpha_s^{(f)}(Q^2)\right]^3
\end{equation}
numerically \cite{ref3} using $\alpha_s^{(5)}(M_Z^2)=0.114$, rather
than using the more conventional approximate solution
\begin{equation}
\frac{\alpha_s^{(f)}(Q^2)}{4\pi} \simeq \frac{1}
         {\beta_0^{(f)}\ln(Q^2/\Lambda^2)}
  -\frac{\beta_1^{(f)}}{(\beta_0^{(f)})^3} 
    \frac{\ln\ln(Q^2/\Lambda^2)}{[\ln (Q^2/\Lambda^2)]^2}
\end{equation}
which becomes sufficiently accurate only for 
$Q^2$ \raisebox{-0.1cm}{$\stackrel{>}{\sim}$} $m_c^2\simeq 2$
GeV$^2$ with \cite{ref3} 
$\Lambda_{\overline{\rm MS}}^{(f=4,5,6)}=257, \, 173.4,\,  68.1$ MeV, 
whereas in LO ($\beta_1\equiv0$) $\Lambda_{\rm LO}^{(4,5,6)} =
175, 132, 66.5$ MeV.  Furthermore, $\beta_0^{(f)}=11-2f/3$ and
$\beta_1^{(f)}=102-38f/3$.  For the $\alpha_s$ matchings at the
$\overline{\rm MS}$ `thresholds' $Q\equiv Q_f = m_f$, i.e.\
$\alpha_s^{(f+1)}(m_{f+1}^2)=\alpha_s^{(f)}(m_{f+1}^2)$, we have
used \cite{ref3} $m_c=1.4$ GeV, $m_b=4.5$ GeV and $m_t=175$ GeV.
On the other hand, we fix $f=3$ in the splitting functions 
$\Delta P_{ij}^{(0,1)}$ in (2.17) and (2.18) for consistency
since we treat the heavy quark sector ($c,b,\ldots$) by the 
perturbatively stable full production cross sections in fixed--order
perturbation theory, i.e. $\gamma^*(Q^2)\gamma(P^2)\to c\bar{c}$
and $\gamma^*(Q^2)g^{\gamma(P^2)} \to c\bar{c}$, etc., keeping
$m_c\neq 0$ as will be discussed below.

For the flavor--nonsinglet case the (matrix) solutions in eqs.\
(2.15)--(2.18) reduce to simple equations for $\Delta
\Sigma^{\gamma(P^2)}\to \Delta q_{\rm NS}^{\gamma(P^2)}$
with $\Delta\vec{k}\,^n \to \Delta k_{\rm NS}^n$ \cite{ref7} and
$\Delta\hat{U}\to\Delta U_{\rm NS}$ expressed in terms of 
$\Delta P_{\rm NS}^{(0)n}$ and $\Delta P_{\rm NS}^{(1)n}$
\cite{ref6,ref16}.

The LO results are of course entailed in the above expressions by
simply dropping all the obvious higher order terms ($\beta_1,\, 
\Delta k^{(1)n},\, \Delta U$).

We shall also compare our quantitative results with the ones based
on the virtual box input (2.13) as suggested in \cite{ref11,ref12}
for $\Lambda^2\ll P^2\ll Q^2$.  Therefore it is also useful to recall
the expression for the $n$--moment, as defined in (2.14), of the
`box' $\Delta q_{\rm had,NLO}^{\gamma(P^2)}(x,P^2)$ in (2.13):
\begin{equation}
\Delta q_{\rm had,NLO}^{\gamma(P^2),n} = 3\, e_q^2\,
    \frac{\alpha}{2\pi}
  \left[ \frac{2}{n}-\frac{4}{n+1}
          -\frac{2}{n^2}+\frac{4}{(n+1)^2}\right]\, .      
\end{equation}

All the above solutions and expressions in Mellin $n$--moment space
can be easily converted into the desired Bjorken--$x$ space by
utilizing a numerical Mellin--inversion as {\mbox described,} 
for example,
in ref.\ \cite{ref6}.

Finally, the heavy quark ($h = c,b,\ldots $) contribution
$g_{1,h}^{\gamma(P^2)}$
to $g_1^{\gamma(P^2)}(x,Q^2)$, as mentioned at the beginning, 
consists of two contributions, the `direct' one and a  `resolved'
one.  The  `direct' contribution derives \cite{ref17} from the 
polarized box diagram $\gamma^*(Q^2)\gamma\to h\bar{h}$, where
the polarized virtual target photon has to be treated as a real
polarized photon $\gamma\equiv\gamma(P^2=0)$ in order to comply
with our continuity condition at $P^2=0$,
\begin{equation}
g_{1,h}^{\rm dir}(x,Q^2) = 3\, e_h^4\,  \frac{\alpha}{2\pi}\,\theta(\beta^2) 
 \left[ (2x-1)\ln\frac{1+\beta}{1-\beta} +\beta(3-4x)\right]
\end{equation}
where $\beta^2=1-4m_h^2/W^2=1-4m_h^2x/(1-x)Q^2$ and 
$h=c,b,t$.  The  `resolved' contribution derives from the polarized
subprocess $\gamma^*(Q^2)g\to h\bar{h}$ and is given by 
\cite{ref18,ref19}
\begin{equation}
g_{1,h}^{\rm res}(x,Q^2) = \int_{y_{\rm min}}^1\frac{dy}{y}
 \Delta g^{\gamma(P^2)}(y,\mu_F^2) 
    \hat{g}_{1,h}^{\gamma^*g\to h\bar{h}}\left(\frac{x}{y},Q^2\right)   
\end{equation}
where $\hat{g}_{1,h}^{\gamma^*g\to h\bar{h}}(x,Q^2)$ is given by
(2.22) with $e_h^4\alpha\to e_h^2\alpha_s(\mu_F^2)/6$,
$y_{\rm min} = x(1+4m_h^2/Q^2)$ and $\mu_F^2\simeq 4m_h^2$.
These contributions add up to $g_{1,h}^{\gamma(P^2)} = 
g_{1,h}^{\rm dir} + g_{1,h}^{\rm res}$.  We state these LO results
for completeness despite the fact that the NLO corrections have 
not yet been calculated and that the heavy quark (charm) contribution
will be immaterial for our more illustrative quantitative purposes.

\setcounter{equation}{0}
\renewcommand{\theequation}{\arabic{section}.\arabic{equation}}
\section{Quantitative Results}

Typical LO and NLO maximal and minimal expectations for 
$\Delta u^{\gamma(P^2)}(x,Q^2)$ and $\Delta g^{\gamma(P^2)}(x,Q^2)$
for real and virtual polarized photons at $Q^2=10$ GeV$^2$ are shown in 
figs.\ 1 and 2 which follow from our   `maximal' input scenario
in (2.10) and the  `minimal' input scenario in (2.11) which in
LO is identical to the  `pointlike' solution in (2.8) as given
by eqs.\ (2.15) and (2.17).  Due to the hadronic component (2.10)
in (2.8), the difference between the `maximal' and  `minimal'
scenario is of course very large for medium and small values
of $x$ for a real $(P^2=0)$ photon, whereas this difference almost
disappears already for $P^2=1$ GeV$^2$, except for 
$\Delta g^{\gamma(P^2)}$ at very small $x$ in fig.\ 2, due to the
suppression of the hadronic contribution by the dipole factor 
$\eta(P^2)$ in (2.10) and (2.11).  Also noteworthy is the 
perturbative LO/NLO stability of $\Delta f^{\gamma(P^2)}(x,Q^2)$ 
which seems to hold in the large and small $x$--region as 
exemplified in figs.\ 1 and 2.  Thus, for 
$P^2$ \raisebox{-0.1cm}{$\stackrel{>}{\sim}$} 1 GeV$^2$,
the structure functions of longitudinally polarized virtual
photons are expected to be dominated by the perturbatively uniquely
calculable  `pointlike' contribution.  The predicted 
$Q^2$--dependence at various fixed values of the virtuality $P^2$
is depicted in figs.\ 3 and 4.  The resulting polarized structure
function $g_1^{\gamma(P^2)}(x,Q^2)$ is shown in fig.\ 5 at some
typical scales $Q^2$ and virtualities $P^2$, with the kinematical
constraint $x\leq Q^2/(Q^2+P^2)$ taken into account.  For 
illustration we also display the  `direct' heavy quark (charm)
contribution according to eq.\ (2.22), whereas the  `resolved'
contribution in (2.23) is much smaller at the scales considered.

We also compared our quantitative results with the ones based 
on the virtual box input (2.13), or (2.21), as studied by Sasaki
and Uematsu \cite{ref11,ref12} for $P^2\gg\Lambda^2$.  Although 
we fully confirm quantitatively their NLO results \cite{ref12}
for $\Delta q^{\gamma(P^2)}(x,Q^2)$ and $\Delta g^{\gamma(P^2)}
(x,Q^2)$, we disagree even with their corrected ones \cite{ref12}
for $g_{1,\ell}^{\gamma(P^2)}(x,Q^2)$, despite the fact that we 
agree with their analytic expressions for $g_{1,\ell}^{\gamma(P^2)}(x,Q^2)$
as given, for example, by
eq.\ (3.16) of ref.\ \cite{ref11}.  This discrepancy is illustrated
in fig.\ 6 where, following \cite{ref11,ref12}, the $Q^2$--evolution
has been performed for fixed $f=3$ flavors, using $\Lambda = 0.2$
GeV.  (Notice that there is a trivial overall normalization 
difference due to the common factor of $\frac{1}{2}$ on the r.h.s.\
of (2.2) which has not been adopted in \cite{ref11,ref12}).

Next we turn to a comparison of our polarized structure functions
with the rather well established unpolarized ones of real as well
as of virtual photons.  The fundamental positivity constraint
$|\Delta\sigma|\leq \sigma$ always refers to the experimentally
measurable cross sections or, in other words, to the directly 
measurable structure functions, i.e.
\begin{equation}
|g_1^{\gamma(P^2)}(x,Q^2)| \leq F_1^{\gamma(P^2)}(x,Q^2)\, ,
\end{equation}
implying $|A_1^{\gamma(P^2)}| \equiv |g_1^{\gamma(P^2)}
/F_1^{\gamma(P^2)}|\leq 1$.  Here $F_1^{\gamma(P^2)}$ 
({\underline{not}} $F_2^{\gamma(P^2)}$) is the spin--averaged
analog of the spin--dependent $g_1^{\gamma(P^2)}$ in (2.2):
\begin{eqnarray}
F_{1,\ell}^{\gamma(P^2)}(x,Q^2) & = & 
 \frac{1}{2} \sum_{q=u,d,s} e_q^2 \left\{ q^{\gamma(P^2)}(x,Q^2)
  + \bar{q}\,^{\gamma(P^2)}(x,Q^2)+\frac{\alpha_s(Q^2)}{2\pi}
    \right.\nonumber\\
& & \left.  
\times  \left[ C_{q,1} \otimes (q+\bar{q})^{\gamma(P^2)} +
      2C_{g,1} \otimes g^{\gamma(P^2)}\right]
        + 2\,e_q^2\,\frac{\alpha}{2\pi}\, C_{\gamma,1}(x)\right\}
\end{eqnarray}
with $q=q_++q_-$ and $g=g_++g_-$ as compared to the 
spin--dependent $\Delta q=q_+-q_-$ and $\Delta g=g_+-g_-$ in
(2.2) in terms of the positive and negative helicity densities
$q_{\pm}$ and $g_{\pm}$.  The NLO coefficient functions in 
(3.2) refer, as in (2.2), to the $\overline{\rm MS}$ 
factorization scheme and are given by
\begin{eqnarray}
C_{q,1}(x) & = & C_{q,2}(x) - \,\frac{4}{3}\,2x\nonumber\\
& &
= \frac{4}{3}\left[(1+x^2)\left( \frac{\ln(1-x)}{1-x}\right)_+
   -\,\frac{3}{2}\, \frac{1}{(1-x)_+}\, 
     -\frac{1+x^2}{1-x}\, \ln x + 3\right.\nonumber\\
& & \left.
   - \left(\frac{9}{2} +\,\frac{\pi^2}{3}\right)
     \delta(1-x)\right]\nonumber\\
C_{g,1}(x) & = & C_{g,2}(x) -\,\frac{1}{2}\, 4x(1-x)\nonumber\\
& &
= \frac{1}{2}\left\{ \left[ x^2+(1-x)^2\right]\,\ln\,\frac{1-x}{x}\,
   +4x (1-x)-1\right\}\nonumber\\
C_{\gamma,1}(x) & = & \frac{3}{(1/2)}\,C_{g,1}(x)
\end{eqnarray}
which have been used in (2.11).  The DIS$_{\gamma,1}$
factorization scheme \cite{ref10} associated with 
$F_{1,\ell}^{\gamma(P^2)}$ and used in the previous Section
is then obtained by absorbing again the entire  `direct'
$C_{\gamma,1}$ term in (3.2) into the $\overline{\rm MS}$ quark
densities $q^{\gamma(P^2)}=\bar{q}\,^{\gamma(P^2)}$:
\begin{eqnarray}
(q+\bar{q})_{{\rm DIS}_{\gamma,1}}^{\gamma(P^2)} & = & 
 (q+\bar{q})^{\gamma(P^2)} + e_q^2\,\frac{\alpha}{\pi}\, C_{\gamma,1}(x)
\nonumber\\
g_{\rm DIS_{\gamma,1}}^{\gamma(P^2)} & = & g^{\gamma(P^2)}\, ,
\end{eqnarray}
in complete analogy to the definition of the polarized 
DIS$_{\Delta\gamma}$ factorization scheme in (2.5).  Again, this
redefinition of parton distributions implies that the unpolarized
NLO($\overline{\rm MS}$) splitting functions $k_{q,g}^{(1)}(x)$
of the photon into quarks and gluons, appearing in the inhomogeneous
NLO $Q^2$--evolution equations \cite{ref6} for 
$f^{\gamma(P^2)}(x,Q^2)$, have to be transformed according to 
\cite{ref6,ref8}
\begin{eqnarray}
k_q^{(1)}|_{\rm DIS_{\gamma,1}} & = & k_q^{(1)} - e_q^2
     P_{qq}^{(0)}\otimes C_{\gamma,1}\nonumber\\
k_g^{(1)}|_{\rm DIS_{\gamma,1}} & = & k_g^{(1)} -2\sum_q
    e_q^2 P_{gq}^{(0)} \otimes C_{\gamma,1}\, 
\end{eqnarray}
similarly to (2.6), where the $k_{q,g}^{(1)}(x)$ can be found, 
for example, in \cite{ref8,ref10}, and $P_{qq}^{(0)} = \Delta
P_{qq}^{(0)}$ and $P_{gq}^{(0)}=\frac{4}{3}\left[1+(1-x)^2\right]/x$.
The NLO expression for $F_{1,\ell}^{\gamma(P^2)}$ in the 
DIS$_{\gamma,1}$ scheme is thus given by (3.2) with $C_{\gamma,1}$
being {\underline{dropped}}.  The relevant DIS$_{\gamma,1}$ parton
distributions are obtained via (2.12) from the ones in the  
DIS$_{\gamma}$ scheme
\cite{ref1} as derived from an analysis of the data on $F_2^{\gamma}
(x,Q^2)$.  The results at $Q^2=10$ GeV$^2$ are shown in fig.\ 7
for a real photon ($P^2=0$) and a virtual one with $P^2=1$ GeV$^2$,
and are compared with the polarized structure function 
$g_{1,\ell}^{\gamma(P^2)}$ for our  `maximal' and  `minimal'
scenario.  For both cases these NLO results are in agreement
\cite{ref10} with the positivity constraint (3.1) which, moreover,
is trivially satisfied in LO \cite{ref10}.  This is also illustrated
in fig.\ 8 where, for completeness, we present the asymmetry
$A_1^{\gamma(P^2)}\equiv g_{1,\ell}^{\gamma(P^2)}/
F_{1,\ell}^{\gamma(P^2)}$ in LO and NLO.

The corresponding asymmetries for the (un)polarized parton
distributions, $A_f^{\gamma(P^2)}\equiv \Delta f^{\gamma(P^2)}/
f^{\gamma(P^2)}$, are depicted in figs.\ 9 and 10 in LO and NLO.
In LO, where cross sections (structure functions) are directly
related to parton densities, the positivity constraint (3.1)
for structure functions implies
\begin{equation}
\Delta f^{\gamma(P^2)}(x,Q^2)|\leq f^{\gamma(P^2)}(x,Q^2)
\end{equation}
which is clearly satisfied, $|A_{u,g}^{\gamma(P^2)}|\leq 1$, as
shown in figs.\ 9 and 10 by the dashed curves.  At NLO, however,
a simple relation between parton distributions and cross sections
no longer holds.  Parton distributions are renormalization and
factorization scheme dependent quantities; although universal,
they are not directly observable, i.e.\ measurable.  Hence there
are NLO contributions which may violate (3.6) in specific cases
\cite{ref10,ref20}.  Such a curiosity occurs for the photonic
parton distributions which, for medium to large values of $x$,
are dominated by the photon's splitting functions $(\Delta)k_{q,g}$
appearing as inhomogeneous terms in the RG $Q^2$--evolution
equations \cite{ref6,ref7,ref8}.  Up to NLO they are given by
\begin{equation}
(\Delta) k_i(x,Q^2) = \frac{\alpha}{2\pi}(\Delta)k_i^{(0)}(x)
  + \frac{\alpha\alpha_s(Q^2)}{(2\pi)^2}\,(\Delta)k_i^{(1)}(x)
\end{equation}
where in LO $(\Delta)k_q^{(0)}=\frac{1}{2}\, 3\, e_q^2\, 2\left[ x^2
\, ^{\hspace{0.13cm} +}_{(-)}
 (1-x)^2\right]$, $(\Delta)k_g^{(0)}=0$ and
the NLO polarized (two--loop) $\Delta k_{q,g}^{(1)}$ are given
in (2.7) and the unpolarized $k_{q,g}^{(1)}$ are as in (3.5).
Our NLO results for $\Delta u^{\gamma(P^2)}$ and 
$\Delta d^{\gamma(P^2)}$ still satisfy the positivity constraint
(3.6) as demonstrated by the solid curves for $A_u^{\gamma(P^2)}$
in fig.\ 9 since in LO $|\Delta k_q^{(0)}|\leq k_q^{(0)}$ despite
the fact that the subleading NLO contributions in (3.7) in
general {\underline{violate}} $|\Delta k_q^{(1)}/k_q^{(1)}|\leq 1$.
The NLO gluon distributions, however, {\underline{violate}}
(3.6) because of the vanishing of the LO terms $(\Delta)k_g^{(0)}=0$
and the now dominant NLO terms $(\Delta)k_g^{(1)}$ in (3.7)
{\underline{violate}} $|\Delta k_g^{(1)}/k_g^{(1)}|\leq 1$.
This violation \cite{ref10} of the NLO gluon `positivity' is
illustrated by the solid curves in fig.\ 10 for $A_g^{\gamma(P^2)}$
where $A_g^{\gamma(P^2)}>1$ for
$x$ \raisebox{-0.1cm}{$\stackrel{>}{\sim}$} 0.6 and 0.7 -- 0.85 
for the `maximal' and   `minimal' scenario, respectively.  
 
Finally it should be mentioned that sum rules for the first $(n=1)$
moment of $g_1^{\gamma(P^2)}$ have been derived for a real $(P^2=0)$
\cite{ref21,ref22} and truly virtual $(P^2\gg \Lambda^2)$
\cite{ref22} polarized photon.  Since our $g_{1,\ell}^{\gamma(P^2)}
(x,Q^2)$ of a virtual photon in (2.2) refers to splitting and
coefficient functions of {\underline{on}}--shell partons and (real)
photons, as dictated by our continuity condition at $P^2=0$, it is
the {\underline{real}}--photon sum rule \cite{ref21,ref22} that 
matters in our case,
\begin{equation}
\int_0^1 dx\, g_{1,\ell}^{\gamma(P^2)}(x,Q^2) = 0\, ,
\end{equation}
which derives from current conservation.  Since this sum rule is
maintained for all $Q^2$ as can be shown by inspecting 
\cite{ref11,ref12} the relevant LO and NLO
evolution kernels and coefficient functions in (2.2) and 
(2.15) -- (2.18)
for $n=1$, in particular the vanishing of the $n=1$ moment of
$\Delta C_{\gamma}$ in (2.4), it can be realized by demanding 
at the input scale $Q^2=\tilde{P}^2$
\begin{equation}
\Delta q^{\gamma(P^2),n=1}(\tilde{P}^2) = 0\, ,
\end{equation}
whereas the photonic gluon distribution remains unconstrained.  Our
LO  `minimal' scenario in (2.11), which corresponds just to the 
`pointlike' solution in (2.8), obviously satisfies (3.9) because
of (2.9).  On the other hand one could rather easily enforce
artificially \cite{ref9} the vanishing of the $n=1$ moment of 
$\Delta q_{\rm had}^{\gamma(P^2)}(x,\tilde{P}^2)$ in the
`maximal' scenario (2.10) as well as of the NLO `minimal' input
in (2.11), but in view of our present complete ignorance of the 
hadronic component of a polarized photon we refrain from doing that.
Since our quantitative speculations refer here mainly to, say,
$x$ \raisebox{-0.1cm}{$\stackrel{>}{\sim}$} $10^{-2}$, the current
conservation constraint (3.9) for the non--vanishing inputs could
be accounted for by contributions from smaller $x$ which do not 
affect of course the evolutions at larger $x$.

For completeness it should also be mentioned that for the truly
virtual region $\Lambda^2\ll P^2\ll Q^2$, i.e.\ if one disregards
the continuity to $P^2=0$, the sum rule (3.8) gets replaced by the
relation \cite{ref22,ref11,ref12}
\begin{equation}
\int_0^1 dx\, g_{1,\ell}^{\gamma(P^2)}(x,Q^2) 
  = 0-\frac{3\alpha}{2\pi} \sum_{q=u,d,s} e_q^4 
     + {\cal{O}}\left(\alpha\alpha_s(Q^2)\right)
\end{equation}
where the LO $-{\cal{O}}(\alpha/\alpha_s)$ contribution vanishes
and the finite NLO$-{\cal{O}}(\alpha)$ term derives essentially 
from the $n=1$ moment of the polarized {\underline{doubly}--virtual
`box' $\gamma^*(Q^2)\gamma(P^2)\to q\bar{q}$ in (2.13) to
$g_{1,\ell}^{\gamma(P^2)}$ in (2.2), cf.\ eq.\ (2.21).  
The NNLO--${\cal{O}}(\alpha\alpha_s)$ contribution in (3.10) has
been calculated as well in \cite{ref22,ref11,ref12}.  (It should
be noted that different normalization and sign conventions for $g_1$
have been used in these latter references.)  Again, this $P^2\gg
\Lambda^2$ approach could be smoothly extrapolated to $P^2=0$, where
the sum rule (3.8) holds, by multiplying the r.h.s.\ of (3.10) by
a form factor like \cite{ref15,ref22} $\,\zeta(P^2)=P^2/(P^2+Q_0^2)$, as
has \mbox{already been} discussed after eq.\ (2.13), where $Q_0^2\simeq 1$
GeV$^2$, i.e.\ tpyically \cite{ref22} 
\mbox{$Q_0^2={\cal{O}}(m_{\rho}^2)$.}
 
\setcounter{equation}{0}
\renewcommand{\theequation}{\arabic{section}.\arabic{equation}}
\section{The Nonresummed QED `Box' Contribution}

An interesting question concerning the photon structure functions
is where the effects due to the RG resummation actually show up.
This question was studied for the unpolarized photon in \cite{ref23}
by comparing the contribution of the non--resummed QED `box' cross
sections for $\gamma^*(Q^2)\gamma(P^2)\to q\bar{q}$ to their QCD
RG--improved counterparts.

In the present context this amounts to comparing $g_{1,\ell}^{\gamma(P^2)}
(x,Q^2)_{\rm box}$, as derived from the longitudinally polarized 
`box' subprocess $\vec{\gamma}\,^*(Q^2)\vec{\gamma}(P^2)\to 
q\bar{q}$, with $g_{1,\ell}^{\gamma(P^2)}(x,Q^2)$ as evaluated 
according to the prescriptions in Section 2.  The general polarized
doubly--virtual box result for the colored three light $q=u,d,s$
quarks ($m_q=0$) is given {\mbox{by \cite{ref17}}}
\begin{eqnarray}
2g_{1,\ell}^{\gamma(P^2)}(x,Q^2)_{\rm box} & = & 
 3\left(\sum_q e_q^4\right)\, \frac{\alpha}{\pi}\,\frac{1}{\bar{\beta}^5}
 \left\{(2x-1)(1-2\delta)\ln\frac{1+\bar{\beta}}{1-\bar{\beta}}
   \right.\\
& & 
  \left. +\bar{\beta}\left[ 2-4x(1-2\delta)-4\delta\right] -8\delta x
   (1-x-\delta)
 \left[ 2\delta x\ln \frac{1+\bar{\beta}}{1-\bar{\beta}}
     -\bar{\beta}\right] \right\}\nonumber
\end{eqnarray}
where $\delta=xP^2/Q^2$ and $\bar{\beta}^2=1-4x\,\delta$.  It should 
be noticed that the third term proportional to $-8\delta x\ldots$ 
does not have any partonic interpretation since it corresponds to
spin--flip transitions for each of the photons with total helicity
conservation, i.e.\ it derives from the combination of helicity 
amplitudes \cite{ref17} $W_{++,00}-W_{0+,-0}$ with $W_{a'b',ab}$
for the transition $ab\to a'b'$.  (We have corrected for a sign
misprint in eq.\ (E.1) of ref.\ \cite{ref17} which results in
$-\bar{\beta}$ in the last term in (4.1)$\,$.)

It is instructive to recall the asymptotic result of our polarized
virtual ($P^2\neq 0$) box expression derived from (4.1) in
the Bjorken limit $Q^2\gg P^2$:
\begin{equation}
2 g_{1,\ell}^{\gamma(P^2)}(x,Q^2)_{\rm box} \simeq 
  3 \left( \sum_q e_q^4\right) \frac{\alpha}{\pi}
     \left\{ (2x-1)\ln \frac{Q^2}{P^2} + (2x-1)
      \left(\ln\frac{1}{x^2} -2\right)\right\}
\end{equation}
where the appropriate  `finite' contribution has been already
used in (2.13).  The {\underline{universal}} process independent
part of this pointlike box expression proportional to 
$\ln Q^2/P^2$ may be used to define formally, as in the case of
an unpolarized photon \cite{ref23,ref24}, light (anti)quark
distributions in the polarized photon $\gamma(P^2)$:
\begin{equation}
2 g_{1,\ell}^{\gamma(P^2)}(x,Q^2)_{\rm box}|_{\rm univ.} \equiv
  \sum_{q=u,d,s} e_q^2\left[ \Delta q_{\rm box}^{\gamma(P^2)}(x,Q^2)
   +\Delta\bar{q}_{\rm box}^{\gamma(P^2)}(x,Q^2)\right]
\end{equation}
with
\begin{equation}
\Delta q_{\rm box}^{\gamma(P^2)}(x,Q^2) = 
 \Delta\bar{q}\,_{\rm box}^{\gamma(P^2)}(x,Q^2) =
   3\, e_q^2\, \frac{\alpha}{2\pi}\, (2x-1) \ln \frac{Q^2}{P^2}\, .
\end{equation}
It should be noted that these naive, i.e.\ not QCD--resummed,
box expressions do not imply a gluon component in the polarized
photon, $\Delta g_{\rm box}^{\gamma(P^2)}(x,Q^2)=0$.

Furthermore, in order to demonstrate the importance of ${\cal{O}}
(P^2/Q^2)$ power corrections in the large $P^2$ region for 
photonic quark distributions, it is sometimes also useful
\cite{ref23} to define, generalizing the definition (4.3), some
`effective' {\underline{non}}--universal (anti)quark distributions  
as common via 
\begin{equation}
2 g_{1,\ell}^{\gamma(P^2)}(x,Q^2)_{\rm box} \equiv
  \sum_{q=u,d,s} e_q^2\left[ \Delta q_{\rm eff}^{\gamma(P^2)}
   (x,Q^2)+\Delta\bar{q}\,_{\rm eff}^{\gamma(P^2)}(x,Q^2)\right]
\end{equation}
where, of course, $\Delta q_{\rm eff}^{\gamma(P^2)} = \Delta
\bar{q}\,_{\rm eff}^{\gamma(P^2)}$ and the full box expression
for $g_{1,{\rm box}}^{\gamma(P^2)}$ is given by (4.1).  The full
box expression implies again $\Delta g_{\rm eff}^{\gamma(P^2)}
(x,Q^2)=0$ in contrast to the QCD resummed finite gluon 
distribution $\Delta g^{\gamma(P^2)}(x,Q^2)$.

Our QCD--resummed total light quark distribution 
$\Delta\Sigma^{\gamma(P^2)}(x,Q^2)\equiv
 2\sum_{q=u,d,s} $ $\Delta q^{\gamma(P^2)}(x,Q^2)$ 
is compared in fig.~11 with
the corresponding universal  `box' expectation according to
(4.4) and with the  `effective' densities as defined in (4.5) which
indicate the relevance of possible ${\cal{O}}(P^2/Q^2)$ terms.
In particular in the small $x$ region,
$x$ \raisebox{-0.1cm}{$\stackrel{<}{\sim}$} $0.3$,
these latter two distributions differ significantly from the
QCD resummed one.  Furthermore the polarized gluon distribution,
which does not exist within the box--approach, becomes
comparable to $\Delta\Sigma^{\gamma(P^2)}$ below 
 $x$ \raisebox{-0.1cm}{$\stackrel{<}{\sim}$} $0.5$ and
dominates, as ususal, in the small--$x$ region, as in the case
of an unpolarized virtual photon \cite{ref23}.  Thus it should
be possible to distinguish between the naive box expectations
and the QCD RG--improved parton distributions of a polarized
photon with future dijet production measurements in polarized
deep inelastic $\vec{e}\vec{p}$ experiments.  Here the 
production rates will, in LO, be related to an effective
polarized parton density \cite{ref25}
\begin{equation}
\Delta\tilde{f}^{\gamma(P^2)}(x,Q^2) = \sum_{q=u,d,s}
 \left[ \Delta q^{\gamma(P^2)}(x,Q^2)+\Delta\bar{q}\,^{\gamma(P^2)}
   (x,Q^2)\right] + \frac{11}{4}\Delta g^{\gamma(P^2)}(x,Q^2)\, ,
\end{equation}
with a similar relation for the proton $\Delta\tilde{f}^p(x,Q^2)$
which is assumed to be known.  This equation is the polarized
counterpart of a similar relation extracted from unpolarized
subprocesses \cite{ref26} as utilized \cite{ref27} for calculating
the high--$p_T$ dijet production rates in unpolarized $ep$ 
collisions.

Finally we compare in fig.\ 12 our QCD RG--improved predictions
for the polarized structure functions $g_{1,\ell}^{\gamma(P^2)}
(x,Q^2)$ for the light $u,d,s$ quarks, to be measured in 
polarized $\vec{e}\,^+\vec{e}\,^-\to e^+e^-X$ experiments, 
with the expectations
of the naive box results in (4.1) and (4.2). Evidently, differences
between
these expectations may be experimentally discernible only in
the small--$x$ region, 
  $x$ \raisebox{-0.1cm}{$\stackrel{<}{\sim}$} $0.2$.

\section{Summary}
The presently unknown parton distributions, $\Delta f^{\gamma(P^2)}
(x,Q^2)$, of the polarized real and virtual photon were studied in
LO and NLO within the context of two extreme scenarios for their
inputs at some low resolution scale.  In particular it was shown
how one may reasonably implement the physical requirement of their
continuity at $P^2=0$ in the nontrivial case of a NLO analysis.
The extreme `maximal' and `minimal' saturation scenarios are 
defined in NLO via eqs.\ (2.10) and (2.11), respectively, where
the choice of the DIS$_{\gamma,1}$ factorization scheme in eq.\ (2.10)
as well as the content of eq.\ (2.11) were dictated by the positivity
constraint $|g_1^{\gamma(P^2)}(x,Q^2)|\leq F_1^{\gamma(P^2)}(x,Q^2)$.
The hadronic input in (2.10) is obtained from an analysis \cite{ref1}
of the unpolarized real photon data on $F_2^{\gamma}(x,Q^2)$
via the relation (2.12).

Finally we compare in figs.\ 11 and 12 the QCD resummed predictions 
of the `maximal' and  `minimal' saturation models with results obtained
within the framework of a simple non--resummed quark  `box' 
$\gamma^*(Q^2)\gamma(P^2)\to q\bar{q}$ calculation (where a
gluon distribution in $\gamma(P^2)$ does not exist), expected to yield
reasonable estimates in the not too small regions of $x$ and $P^2$.

\noindent{\large\bf{Acknowledgements}}

We thank I.\ Schienbein for helpful discussions concerning the
calculation and results of the polarized box contribution.
This work has been supported in part by the `Bundesministerium
f\"ur Bildung, Wissenschaft, Forschung und Technologie', Berlin/Bonn.

\newpage

\noindent{\large{\bf{\underline{Figure Captions}}}}
\begin{itemize}
\item[\bf{Fig.\ 1}.]  Typical LO and NLO(DIS$_{\Delta\gamma}$)
       expectations for the parton densities of a real $(P^2= 0)$
       and virtual polarized photon at a common scale of $Q^2=10$
       GeV$^2$, which follow from our  `maximal' and `minimal' input
       scenarios in eqs.\ (2.10) and (2.11), respectively.

\item[\bf{Fig.\ 2}.]  Same as in fig.\ 1 but plotted for a logarithmic
       $x$--scale in order to illustrate the small--$x$ structure of 
       the polarized distributions.

\item[\bf{Fig.\ 3}.] The predicted $Q^2$--dependence of 
       $\Delta u^{\gamma(P^2)}(x,Q^2)$ in NLO(DIS$_{\Delta\gamma}$) at
       various fixed values of the virtuality $P^2$ according to the
       `maximal' and  `minimal' input scenarios in eqs.\ (2.10) and
       (2.11), respectively.  For $P^2 = 1$ GeV$^2$ the results at 
       $Q^2=2$ GeV$^2$ are, for obvious reasons, not displayed anymore.

\item[\bf{Fig.\ 4}.] As fig.\ 3 but for $\Delta g^{\gamma(P^2)}(x,Q^2)$
       in NLO.

\item[\bf{Fig.\ 5}.] The resulting NLO predictions of the polarized
       structure function $g_{1,\ell}^{\gamma(P^2)}(x,Q^2)$ for the 
       light $u,d,s$ quarks as defined in (2.2), according to the
       `maximal' and `minimal' input scenarios in eqs.\ (2.10) and 
       (2.11), respectively.  Notice that the virtual photon structure
       function is kinematically constrained within $0\leq x \leq
       (1+P^2/Q^2)^{-1}$.  For comparison the charm contribution at
       $Q^2=10$ GeV$^2$ is shown as well according to the  `direct'
       box expression (2.22) using $m_c=1.4$ GeV.  The  `resolved'
       contribution in (2.23) is marginal in the kinematic region
       considered.

\item[\bf{Fig.\ 6}.] The polarized virtual photon structure function
       $g_{1,\ell}^{\gamma(P^2)}(x,Q^2)$ for the $f=3$ light quark flavors
       at $Q^2=30$ GeV$^2$ and $P^2=1$ GeV$^2 \gg \Lambda^2$, where
       $N=(3/2)\sum_q e_q^4(\alpha/\pi)\ln (Q^2/P^2)$ with
       $\sum_q e_q^4=2/9$.  Our NLO result is compared with the one
       presented in the second reference of ref.\ \cite{ref12}.

\item[\bf{Fig.\ 7}.] The unpolarized and polarized structure functions
       $F_{1,\ell}^{\gamma(P^2)}(x,Q^2)$ and $g_{1,\ell}^{\gamma(P^2)}
       (x,Q^2)$  in NLO for the three light $u,d,s$ quarks as defined
       in eqs.\ (3.2) and (2.2), respectively.  $F_{1,\ell}^{\gamma(P^2)}$
       has been calculated according to the analysis of ref.\ \cite{ref1}.
       The results for the polarized structure function 
       $g_{1,\ell}^{\gamma(P^2)}$ refer to the `maximal' and `minimal'
       scenarios in eqs.\ (2.10) and (2.11), respectively.  The 
       Bjorken--$x$ is kinematically constrained by 
       $x\leq (1+P^2/Q^2)^{-1}$.

\item[\bf{Fig.\ 8}.] The spin asymmetries $A_1^{\gamma(P^2)}(x,Q^2)\equiv
       g_{1,\ell}^{\gamma(P^2)}/F_{1,\ell}^{\gamma(P^2)}$ in LO and NLO
       for the  `maximal' and `minimal' scenario for 
       $g_{1,\ell}^{\gamma(P^2)}$.  The NLO results are of course
       directly related to the ones of fig.\ 7.

\item[\bf{Fig.\ 9}.] The up--quark spin asymmetry $A_u^{\gamma(P^2)}(x,Q^2)
       \equiv \Delta u^{\gamma(P^2)}/u^{\gamma(P^2)}$ with 
       $\Delta u^{\gamma(P^2)}(x,Q^2)$ being calculated according to 
       the  `maximal' and `minimal' scenarios in eqs.\ (2.10) and
       (2.11), respectively, and the unpolarized $u^{\gamma(P^2)}(x,Q^2)$
       is calculated according to the DIS${_\gamma}$ results of ref.\ 
       \cite{ref1} using eq.\ (2.12).
       The polarized NLO distributions refer to the DIS$_{\Delta\gamma}$
       factorization scheme defined in (2.5), whereas the unpolarized
       NLO distributions are calculated in the DIS$_{\gamma,1}$ scheme
       as defined in (3.4).

\item[\bf{Fig.\ 10}.] As in fig.\ 9 but for the gluon spin asymmetry
       $A_g^{\gamma(P^2)}(x,Q^2)\equiv
       \Delta g^{\gamma(P^2)}/g^{\gamma(P^2)}$.

\item[\bf{Fig.\ 11}.] Comparing the LO QCD--resummed total polarized 
      light quark distribution $\Delta\Sigma^{\gamma(P^2)}(x,Q^2)\equiv
      2 \sum_{q=u,d,s}\Delta q^{\gamma(P^2)}$ and the polarized gluon
      distribution $\Delta g^{\gamma(P^2)}(x,Q^2)$ in the `maximal' and
      `minimal' scenario with the naive universal `box' results defined
      in (4.4), and with the  `effective' distributions derived from (4.5).  
      Notice that it is the quantity $(11/4)\Delta 
      g^{\gamma(P^2)}$, which appears in the effective polarized parton
      density in (4.6), that will be directly accessible by future
      experiments.

\item[\bf{Fig.\ 12}.] Comparing the LO-- and NLO--QCD results for the
      polarized structure function $g_{1,\ell}^{\gamma(P^2)}(x,Q^2)$ 
      for the light $u,d,s$ quarks for the  `maximal' and `minimal'
      input scenarios in (2.10) and (2.11) with the expectations due
      to the  `full box' in (4.1) and its `asymptotic box' 
      approximation given in (4.2) for $Q^2\gg P^2$.
\end{itemize}
\clearpage
\pagestyle{empty}

\begin{figure}
\begin{center}
\epsfig{file=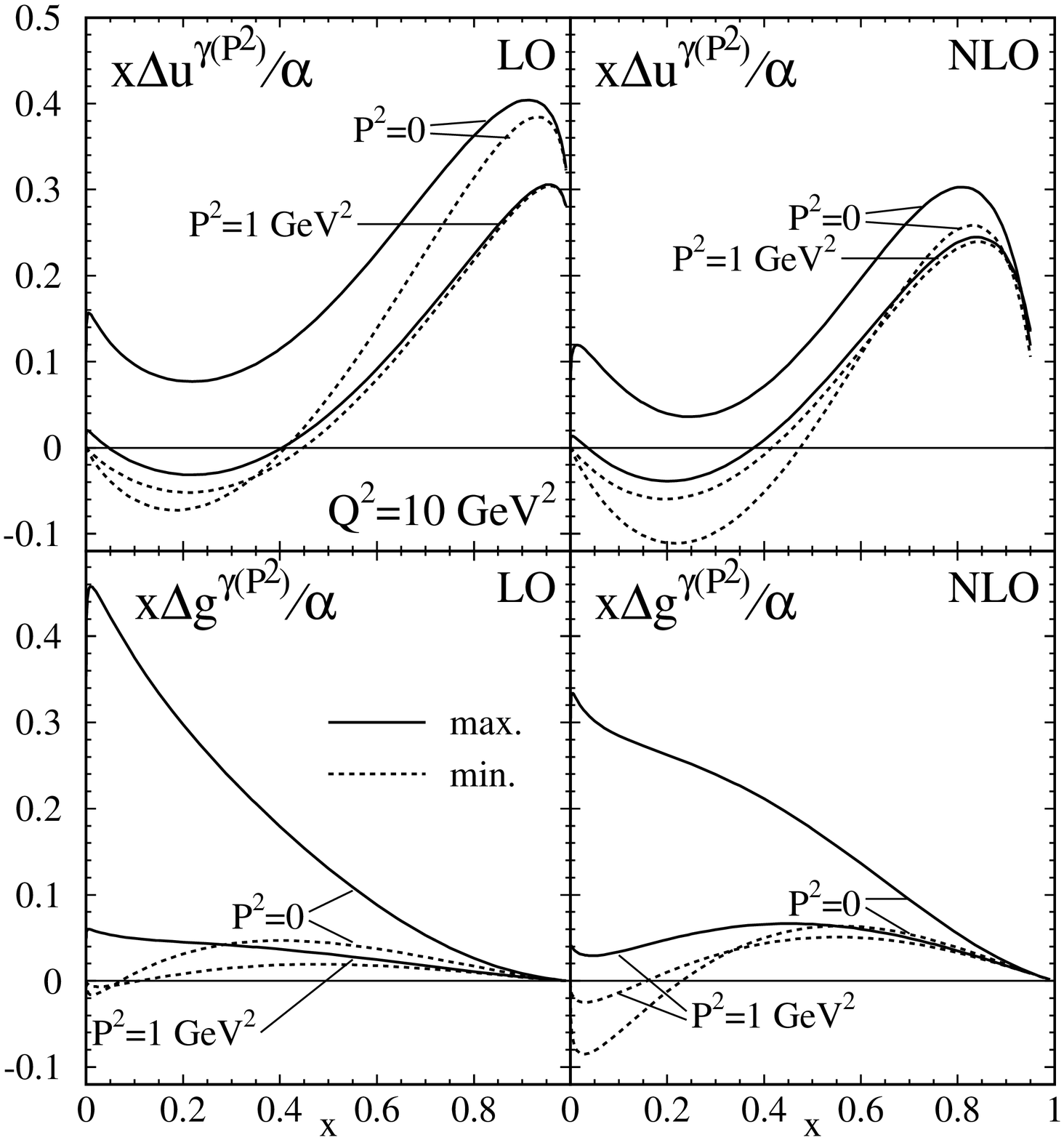,width=\textwidth}
\vspace*{1cm}
\\{\large\bf Fig. 1}
\end{center}
\end{figure}

\begin{figure}
\begin{center}
\epsfig{file=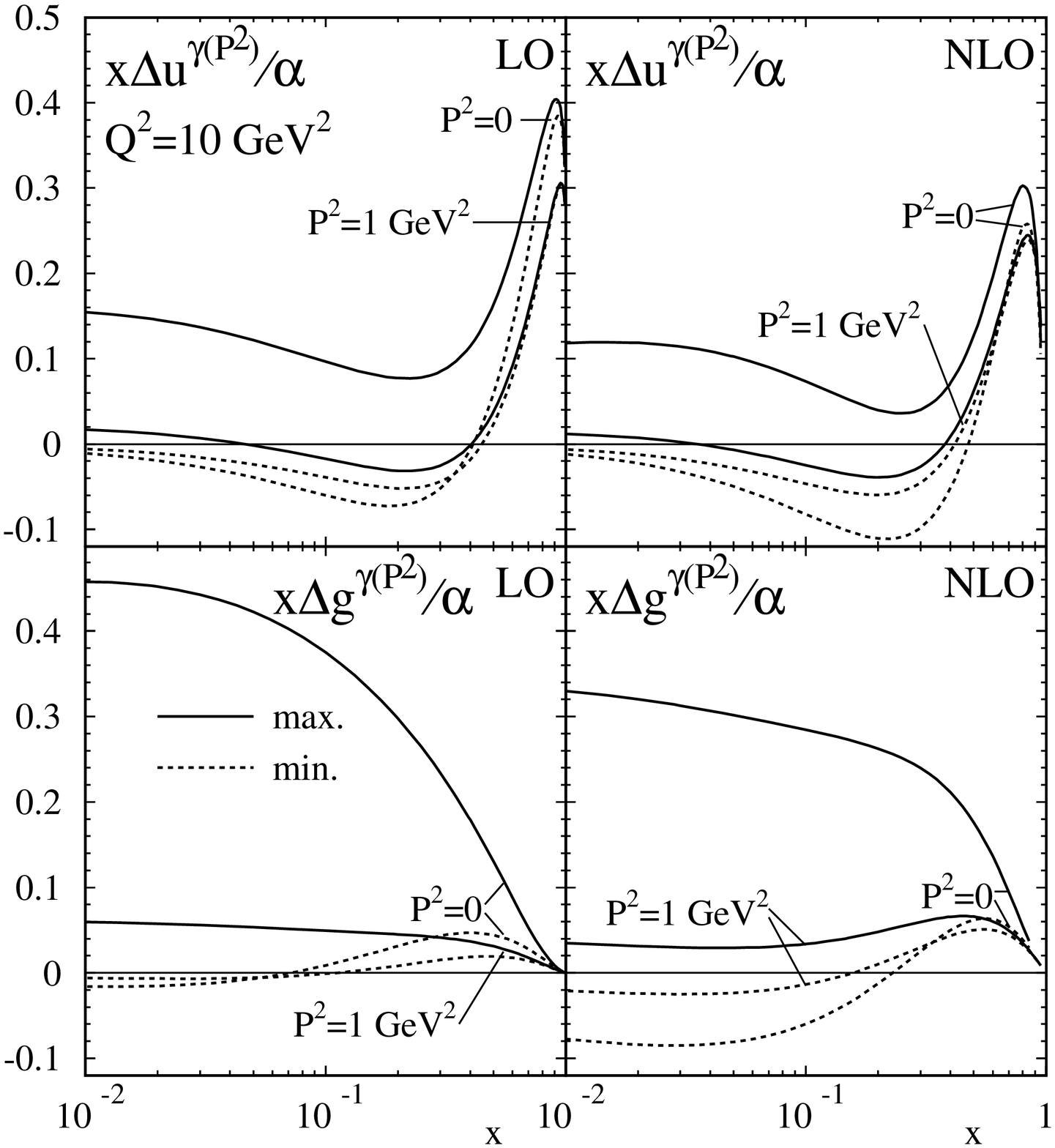,width=\textwidth}
\vspace*{1cm}
\\{\large\bf Fig. 2}
\end{center}
\end{figure}

\begin{figure}
\begin{center}
\epsfig{file=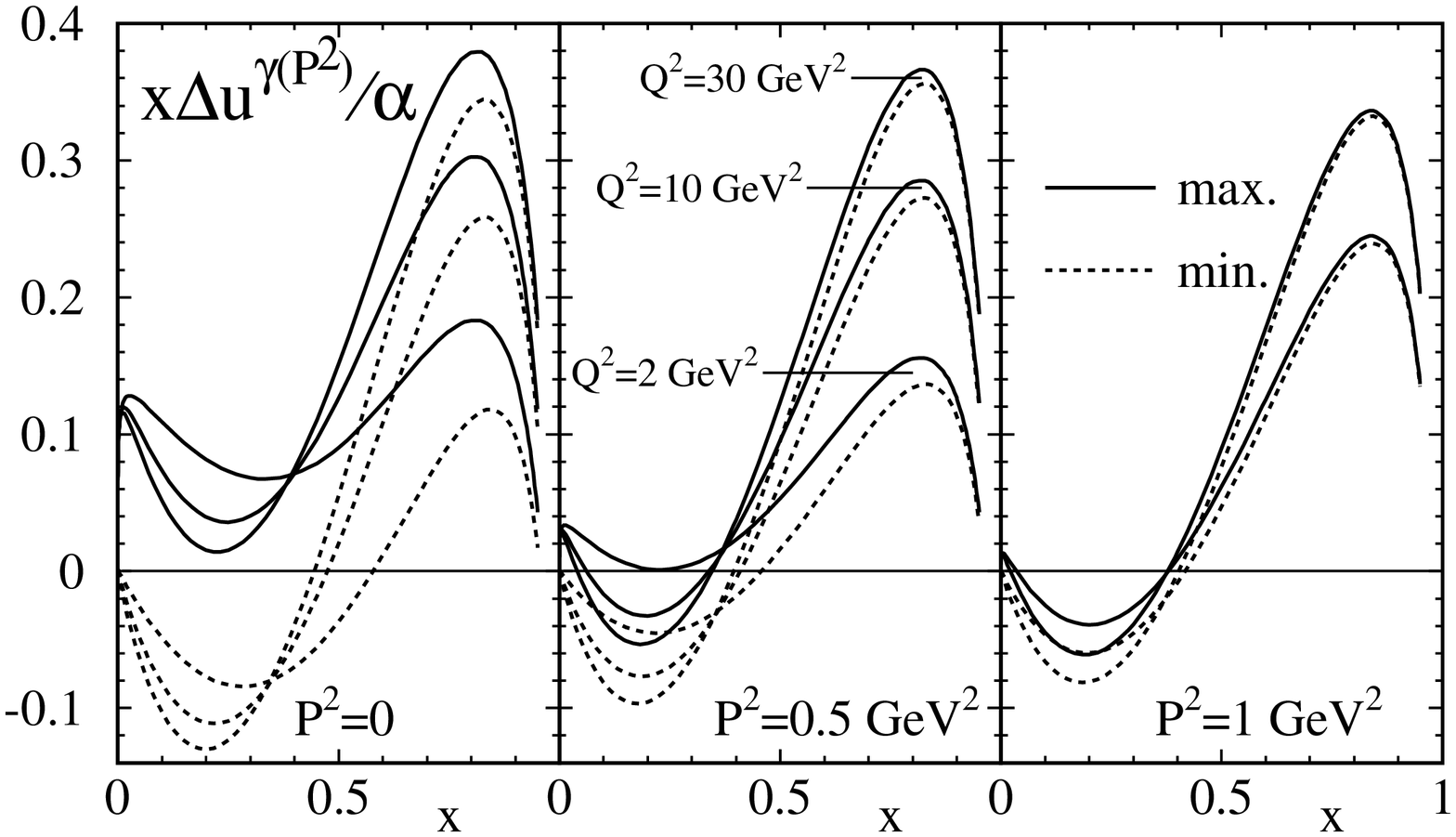,width=\textwidth}
\\{\large\bf Fig. 3}\\
\vspace*{2cm}
\epsfig{file=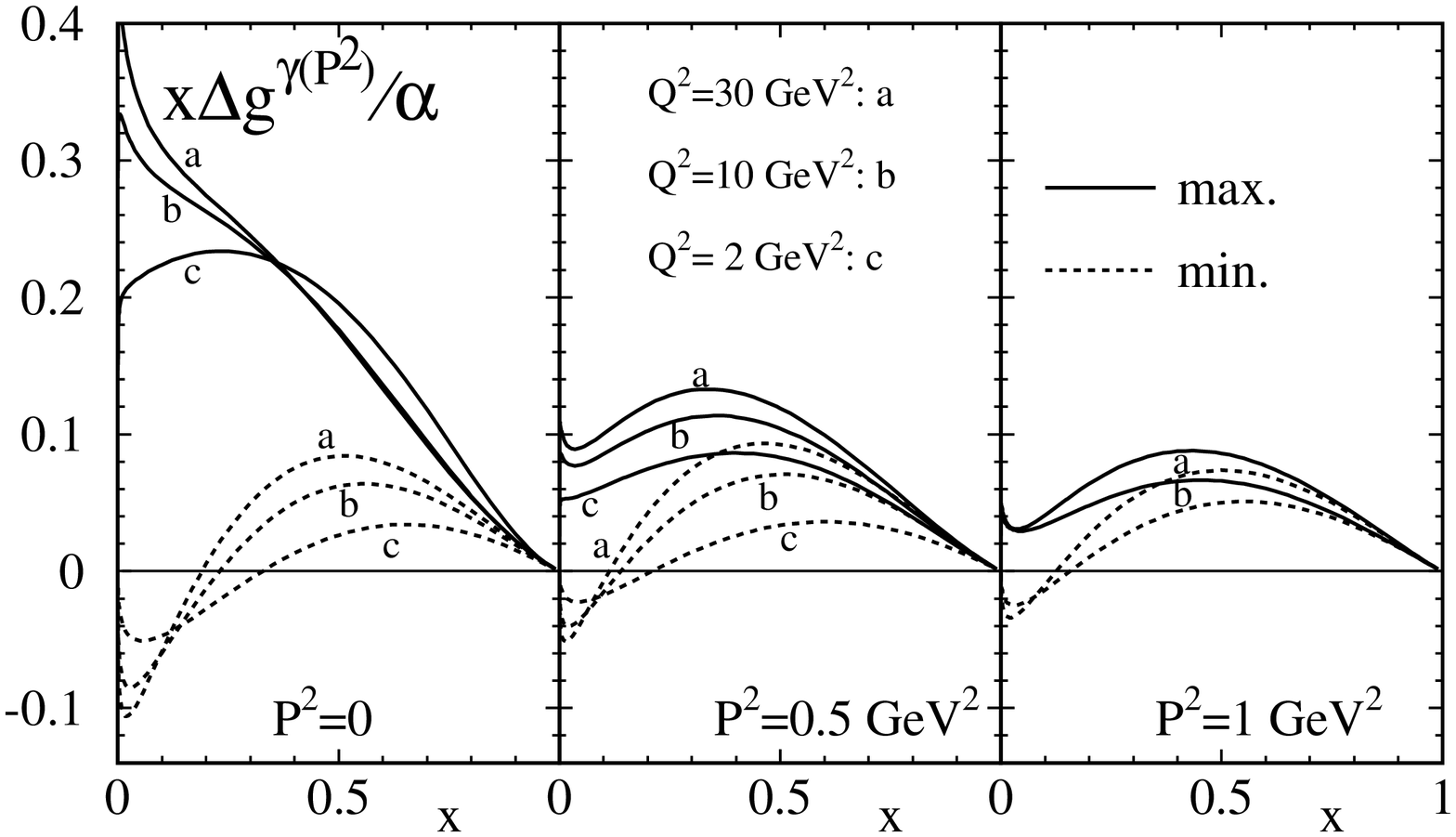,width=\textwidth}
\\{\large\bf Fig. 4}
\end{center}
\end{figure}
\clearpage

\begin{figure}
\begin{center}
\epsfig{file=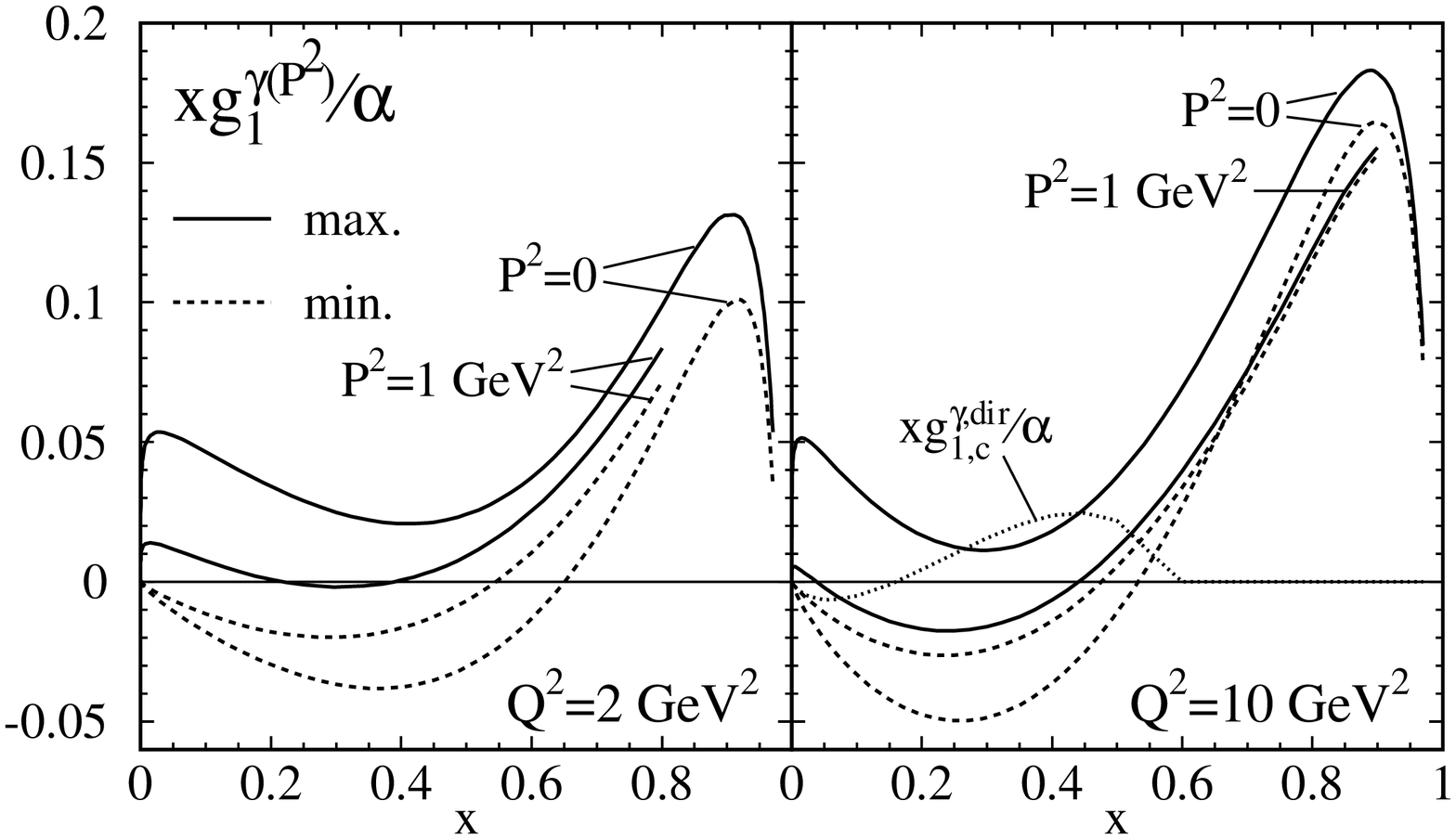,width=\textheight,angle=90}
\put(-5.,280.){\rotatebox{90}{\large\bf Fig. 5}}
\end{center}
\end{figure}

\begin{figure}
\begin{center}
\epsfig{file=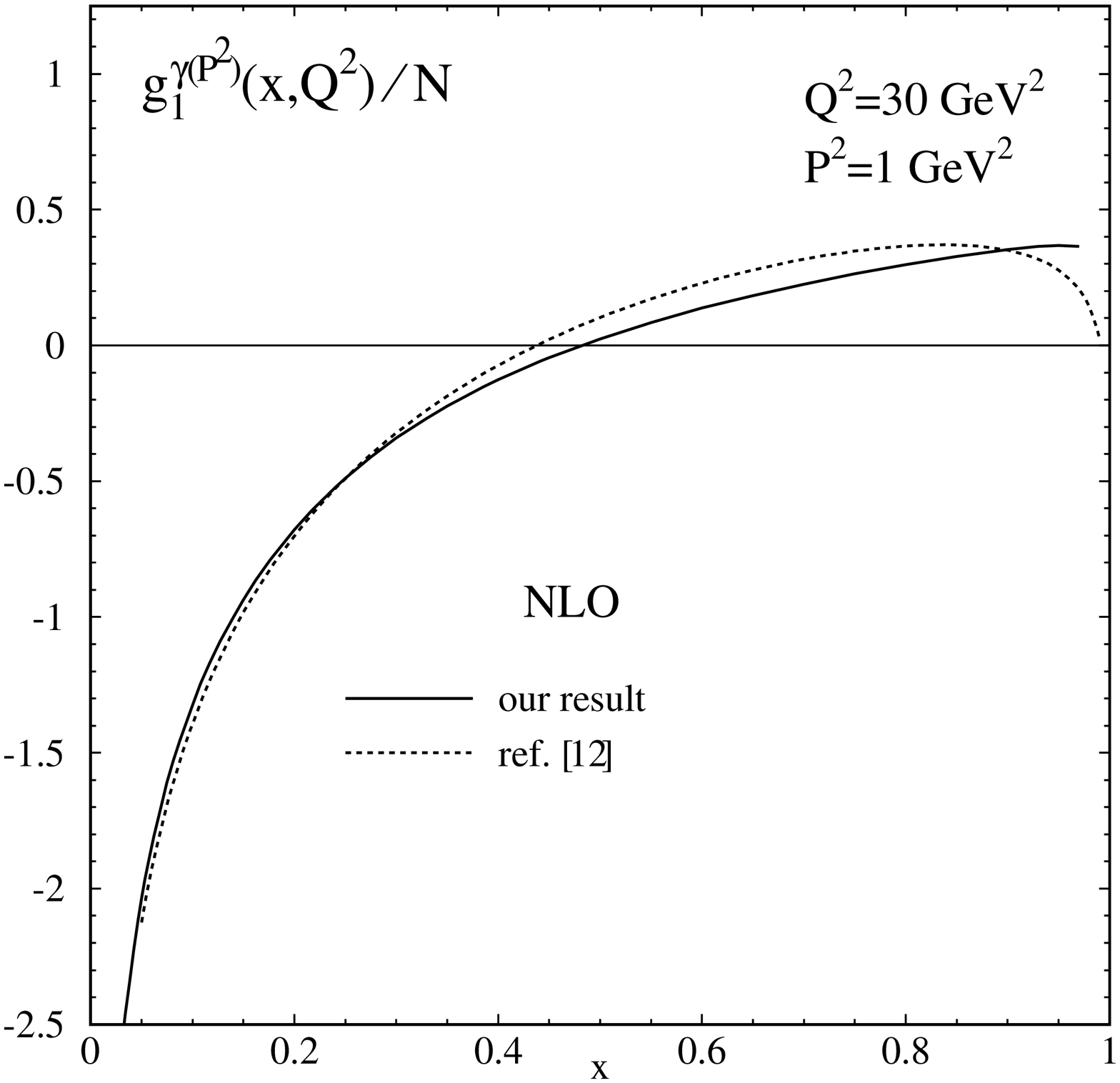,width=\textwidth}
\vspace*{1cm}
\\{\large\bf Fig. 6}
\end{center}
\end{figure}

\begin{figure}
\begin{center}
\epsfig{file=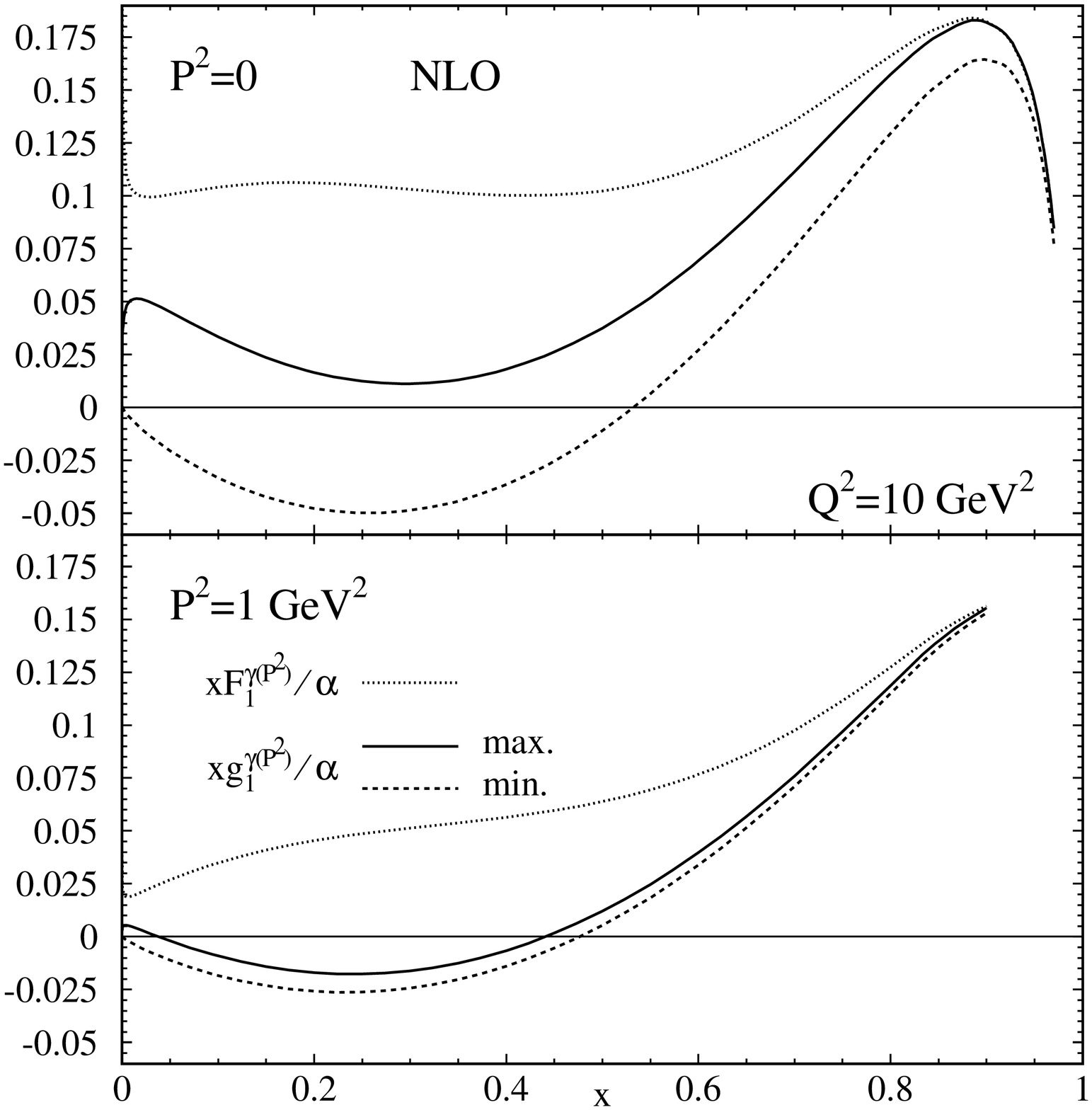,width=\textwidth}
\vspace*{1cm}
\\{\large\bf Fig. 7}
\end{center}
\end{figure}

\begin{figure}
\begin{center}
\epsfig{file=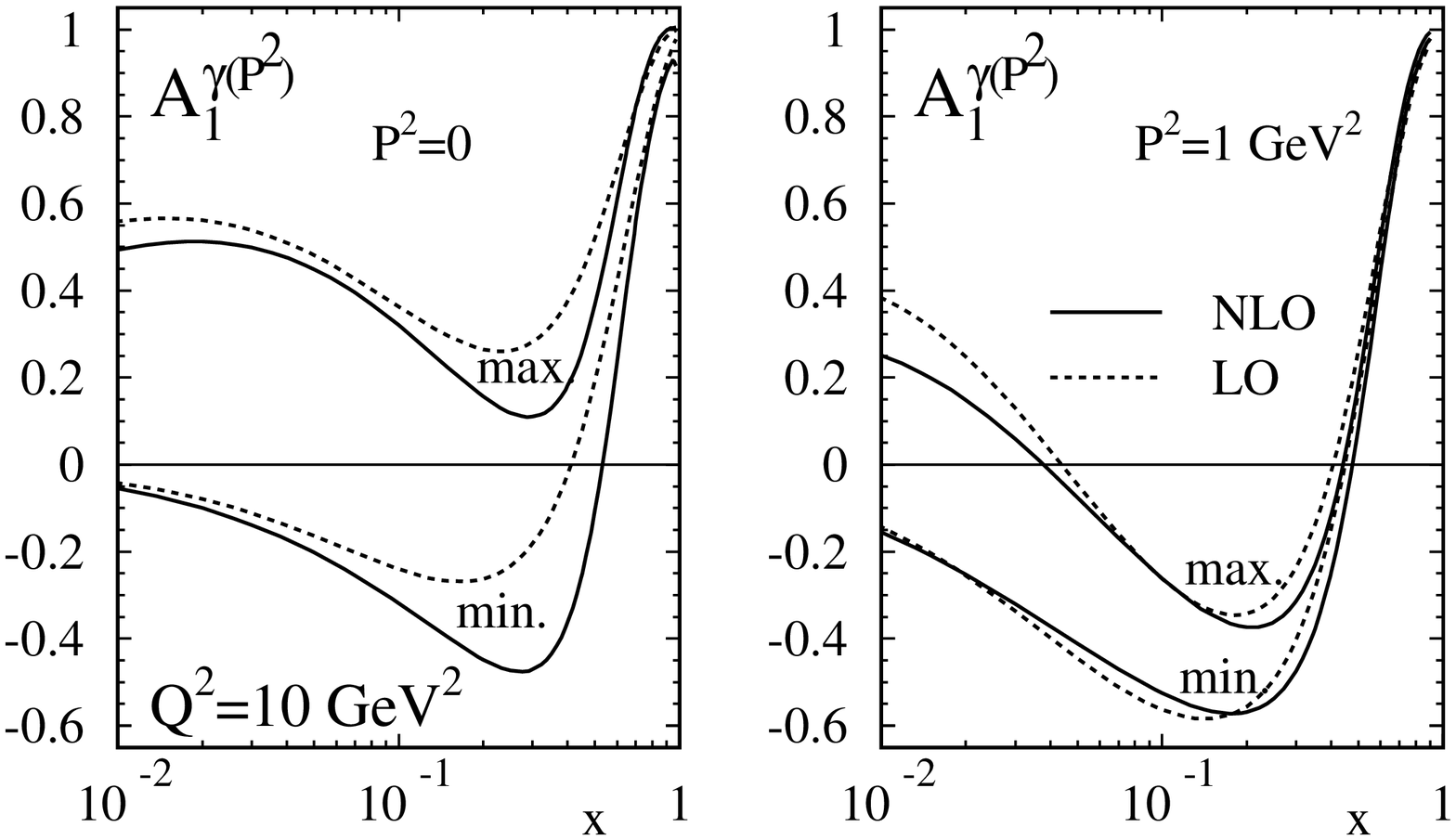,width=\textheight,angle=90}
\put(-5.,280.){\rotatebox{90}{\large\bf Fig. 8}}
\end{center}
\end{figure}

\begin{figure}
\begin{center}
\epsfig{file=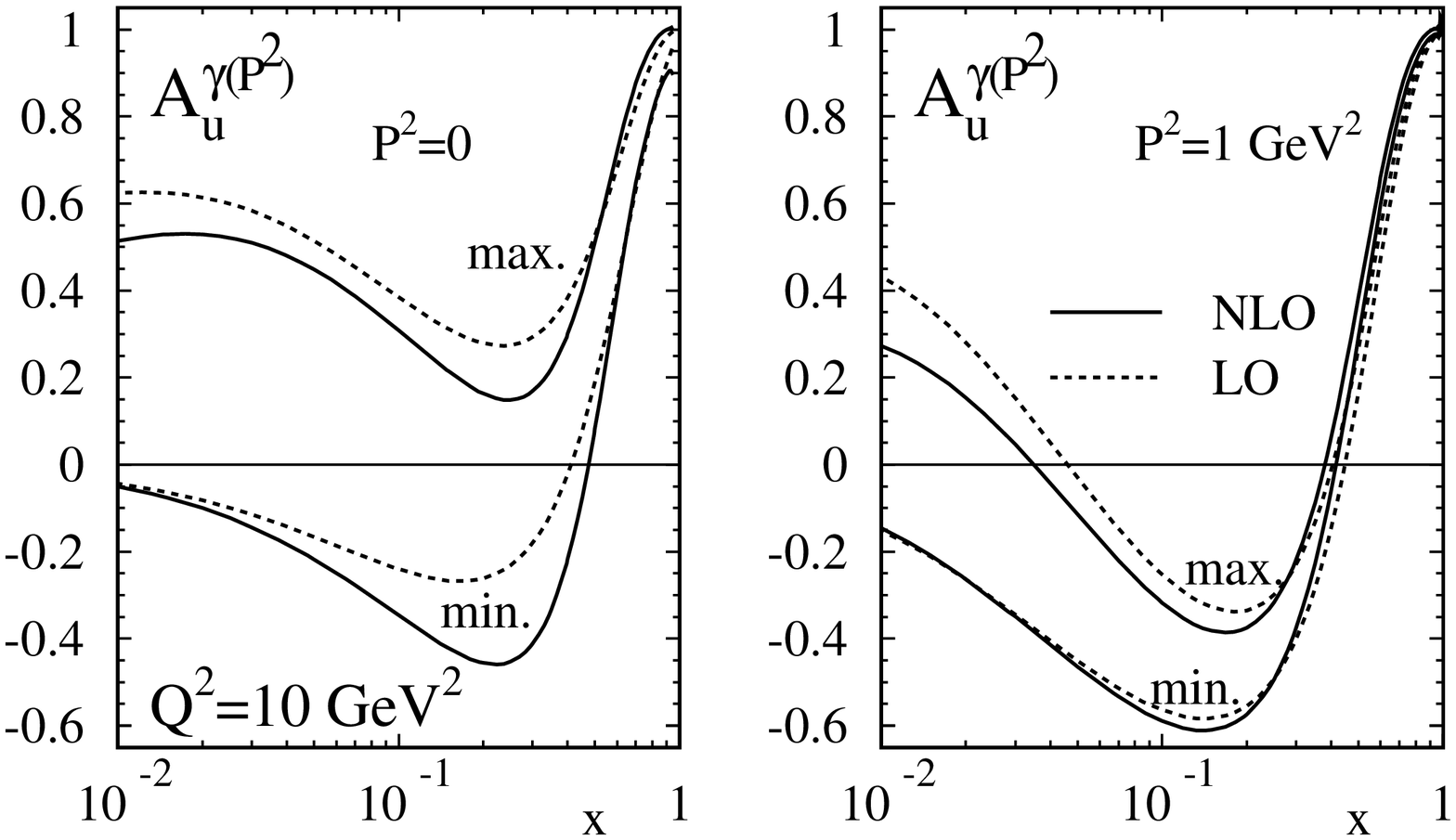,width=\textwidth}
\\{\large\bf Fig. 9}\\
\vspace*{2cm}
\epsfig{file=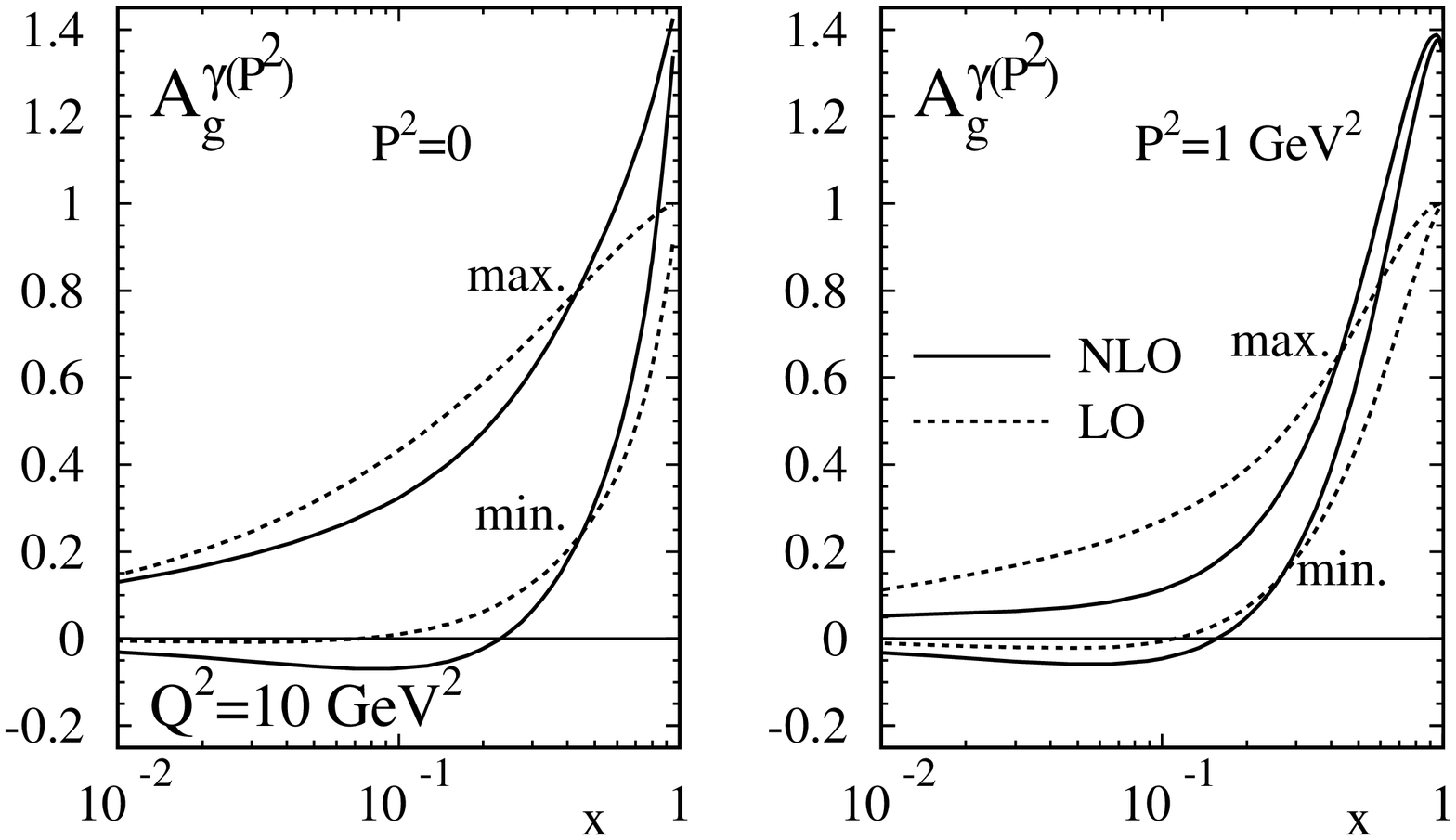,width=\textwidth}
\\{\large\bf Fig. 10}\\
\end{center}
\end{figure}

\begin{figure}
\begin{center}
\epsfig{file=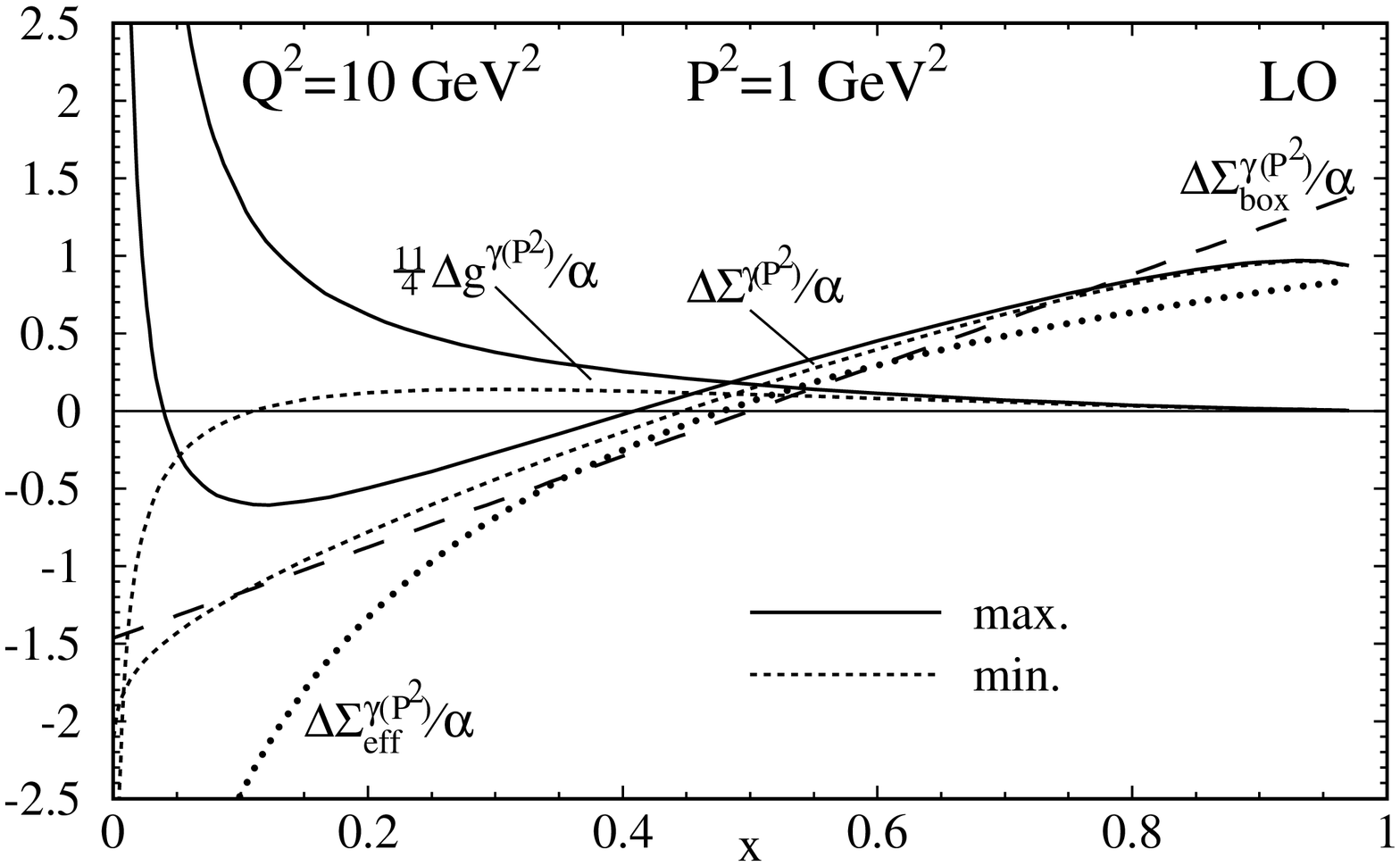,width=\textwidth}
\\{\large\bf Fig. 11}\\
\vspace*{2cm}
\epsfig{file=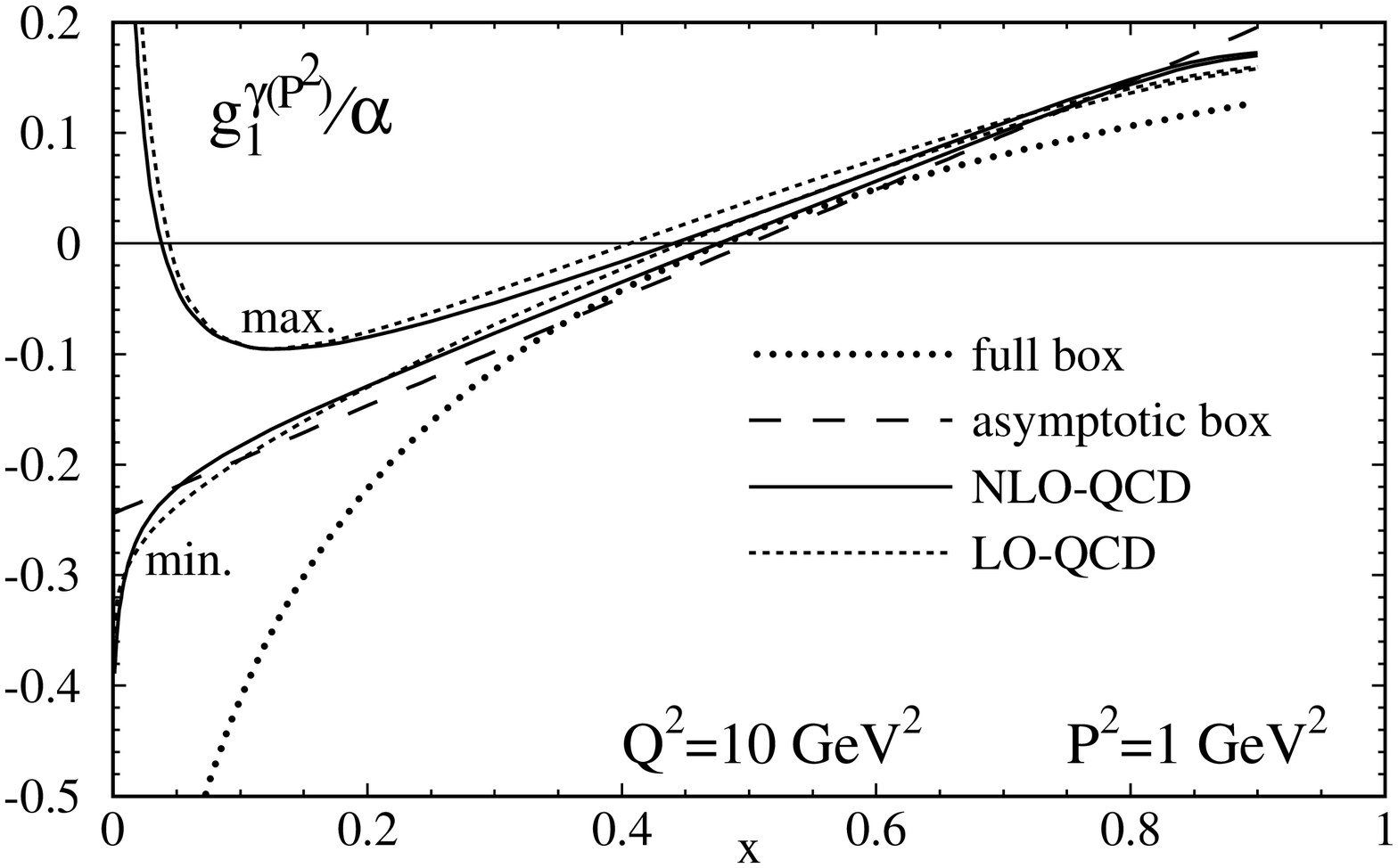,width=\textwidth}
\\{\large\bf Fig. 12}
\end{center}
\end{figure}

\end{document}